\documentclass[a4paper,oneside]{scrartcl}

\newcommand{\Title}{Complexity of Model Checking for Modal Dependence Logic}
\newcommand{\Author}{
Johannes Ebbing%
\thanks{
Leibniz Universit{\"a}t Hannover,
Theoretical Computer Science,
Appelstr.~4, 30167~Hannover, Germany,
\{ebbing,lohmann\}@thi.uni-hannover.de},
Peter Lohmann\footnotemark[1]
}
\newcommand{\PDFAuthor}{Johannes Ebbing, Peter Lohmann}
\newcommand{\Keywords}{dependence logic, modal logic, model checking, computational complexity}
\newcommand{\SubjectClassifiers}{F.2.2 Complexity of proof procedures; F.4.1 Modal logic; D.2.4 Model checking}

\usepackage{amsmath, amssymb}
\usepackage{stmaryrd}
\usepackage{xspace}
\usepackage{xcolor}
\usepackage{needspace}
\usepackage{enumerate}
\usepackage{ifthen}
\usepackage{fancyhdr}
\usepackage{tikz}
  \usetikzlibrary{arrows,automata,positioning}
 \usepackage{listings}
  
  \lstdefinelanguage{pseudo}{
    morekeywords={if,elseif,then,return,end,choose,guess,when,for,foreach,case},
    morekeywords=[3]{false,true,and,or,not},
    morecomment=[l]{//}
  }
\newlength\problemlength
\newcommand\problemdef[3]{%
\begin{list}{}{\topsep \medskipamount \labelwidth\problemlength \labelsep.7em \rightmargin1.5em
\leftmargin\problemlength \advance\leftmargin by3em
\parsep0ex \itemsep.2ex plus.1ex}
\item[{\sl Problem:\hfill}] #1
\item[{\sl Input:  \hfill}] #2
\item[{\sl Output: \hfill}] #3
\end{list}
}
\newlength\dproblength
\newcommand{\comment}[2][]{{\color{blue}\scriptsize+++\ifthenelse{\equal{}{#1}}{}{#1:\ }#2---------}\marginpar{{\color{blue}$\bullet$}}}

\newcommand{\halfliteral}[1]{\protect\ensuremath{#1}}
\newcommand{\literal}[1]{\halfliteral{#1}\xspace}

\newcommand{\numberClassFont}[1]{\mathbb{#1}}     
\newcommand{\complexityClassFont}[1]{\textsl{#1}} 
\newcommand{\logicClFont}[1]{\mathsf{#1}}        
\newcommand{\problemFont}[1]{\mathrm{#1}}         
\newcommand{\mathCommandFont}[1]{\mathrm{#1}}     

\newcommand{\N}{\protect\ensuremath{\numberClassFont{N}}\xspace}

\newcommand{\enc}[1]{\ensuremath\langle#1\rangle}

\newcommand{\PTIME}{\protect\ensuremath{\complexityClassFont{P}}\xspace}
\newcommand{\NP}{\protect\ensuremath{\complexityClassFont{NP}}\xspace}

\newcommand{\NEXPTIME}{\ensuremath{\complexityClassFont{NEXP}}\-\ensuremath{\complexityClassFont{TIME}}\xspace}

\newcommand{\leqpm}{\protect\ensuremath{\leq^\mathCommandFont{p}_\mathCommandFont{m}}}
\newcommand{\equivpm}{\protect\ensuremath{\equiv^\mathCommandFont{p}_\mathCommandFont{m}}}

\newcommand{\false}{\literal{\bot}}
\newcommand{\true}{\literal{\top}}
\newcommand{\aneg}{\protect\ensuremath{\neg}\xspace}
\newcommand{\nor}{\ovee}
\DeclareMathOperator*{\bignor}{\text{{\huge\ensuremath{\varovee}}}}
\newcommand{\sor}{\vee}
\newcommand{\dep}[2][]{\protect\ensuremath{\mathop{=}}\ifthenelse{\equal{#1}{}}{\ifthenelse{\equal{#2}{}}{}{\nolinebreak\ensuremath{(#2)}}}{\nolinebreak\ensuremath{(#1,#2)}}}
  \newcommand{\allMDL}{\AX,\EX,\allowbreak\wedge,\allowbreak\sor,\allowbreak\nor,\allowbreak\neg,\allowbreak\dep{}}

\newcommand{\MDL}[1][]{\ensuremath{\logicClFont{MDL}\ifthenelse{\equal{#1}{}}{}{(\allowbreak#1)}}\xspace}
\newcommand{\CTL}[1][]{\ensuremath{\logicClFont{CTL}\ifthenelse{\equal{#1}{}}{}{(#1)}}\xspace}
\newcommand{\LTL}[1][]{\ensuremath{\logicClFont{LTL}\ifthenelse{\equal{#1}{}}{}{(#1)}}\xspace}
\newcommand{\CTLs}[1][]{\ensuremath{\logicClFont{\CTL^*}\ifthenelse{\equal{#1}{}}{}{(#1)}}\xspace}
\newcommand{\ML}{\ensuremath{\mathsf{ML}}\xspace}

\newcommand{\MDLk}[1][]{\ensuremath{\logicClFont{MDL_k}\ifthenelse{\equal{#1}{}}{}{(\allowbreak#1)}}\xspace}

\newcommand{\MDLMC}[1][]{\ensuremath{\MDL\text{-}}\allowbreak\ensuremath{\problemFont{MC}\ifthenelse{\equal{#1}{}}{}{(#1)}}\xspace}
\newcommand{\MDLMCparas}[3][]{\ensuremath{\problemFont{\MDL\ifthenelse{\equal{#2}{}}{}{_{\mathnormal{#2}}}}\text{-}}\allowbreak\ensuremath{\problemFont{MC\ifthenelse{\equal{#3}{}}{}{_{\mathnormal{#3}}}}\ifthenelse{\equal{#1}{}}{}{(#1)}}\xspace}
\newcommand{\MDLMCpara}[2][]{\MDLMCparas[#1]{#2}{}}
\newcommand{\MDLMCk}[1][]{\MDLMCpara[#1]{k}}

\newcommand{\cnf}{\protect\ensuremath{\problemFont{3CNF}}\xspace}
\newcommand{\CNF}{\cnf}

\newcommand{\ThreeSAT}{\ensuremath{\problemFont{3SAT}}\xspace}

\newcommand{\powerset}[1]{\protect\ensuremath{\mathcal{P}}(#1)\xspace}
\newcommand{\set}[3][]{\protect\ensuremath{\left\{#2\;\middle|\;\ifthenelse{\equal{#1}{}}{\text{#3}}{\parbox{#1}{#3}}\right\}}}

\DeclareMathOperator{\AX}{\Box}
\DeclareMathOperator{\EX}{\Diamond}

\providecommand{\dfn}{\mathrel{\mathop:}=}
\providecommand{\ddfn}{\mathrel{\mathop{{\mathop:}{\mathop:}}}=}
\providecommand{\anfz}[1]{\text{``}#1\text{''}}

\newcommand{\ie}{\allowbreak{}i.\,e.,\ }

\usepackage[
  colorlinks=false,pdfdisplaydoctitle=false,pdfkeywords={\Keywords},pdftitle={\Title},pdfauthor={\PDFAuthor}
]{hyperref}

\newcommand\thmsymbol{\quad\text{\scriptsize\color{gray}\ensuremath{\dashv}}}
\newcommand\claimsymbol{\quad\text{\scriptsize\color{lightgray}\ensuremath{\dashv}}}
\newcommand\dfnsymbol{\quad\text{\Large\color{gray}\ensuremath{\lrcorner}}}
\newcommand\proofsymbol{\quad\text{\scriptsize \ensuremath{\blacksquare}}}
\newcommand\altproofsymbol{\quad\text{\normalsize\ensuremath{\blacktriangleleft}}}

\usepackage[amsmath,hyperref,thmmarks]{ntheorem}
\theoremstyle{plain}
\theorembodyfont{\itshape}
\theoremheaderfont{\normalfont\bfseries}
\theoremseparator{.}
\theoremsymbol{\thmsymbol}
\newtheorem{theorem}{Theorem}[section]
\newtheorem{lemma}[theorem]{Lemma}
\newtheorem{proposition}[theorem]{Proposition}

\theoremsymbol{\claimsymbol}

\theorembodyfont{\upshape}
\theoremsymbol{\dfnsymbol}
\newtheorem{definition}[theorem]{Definition}
\newtheorem{example}[theorem]{Example}

\theoremstyle{nonumberplain}
\theoremheaderfont{\scshape}
\theorembodyfont{\normalfont}
\theoremseparator{.}
\theoremsymbol{\proofsymbol}
\newtheorem{proofcontent}{Proof}
\newenvironment{proof}{\begin{proofcontent}\setcounter{claim}{0}}{\end{proofcontent}\setcounter{claim}{0}}
\newtheorem{proofsketchcontent}{Proof (Sketch)}

\qedsymbol{\proofsymbol}
\theoremstyle{empty}
\theoremheaderfont{\scshape}
\theorembodyfont{\normalfont}
\theoremseparator{.}
\theoremsymbol{\altproofsymbol}
\newtheorem{namedproofcontent}{}

\usepackage[capitalize]{cleveref}
\Crefname{algorithm}{Algorithm}{Algorithms}
\Crefname{listing}{Algorithm}{Algorithms}
\Crefname{figure}{Figure}{Figures}
\Crefname{note}{Note}{Notes}
\Crefname{claim}{Claim}{Claims}

\begin{document}

\title{\Title}

\author{\Author}

\maketitle

\thispagestyle{fancy}
\renewcommand{\headrulewidth}{0pt}
\renewcommand{\footrulewidth}{0pt}
\fancyhf{}
\fancyfoot[L]{
\scriptsize
\begin{tabular}[c]{@{}ll@{}}
  \includegraphics[height=5ex]{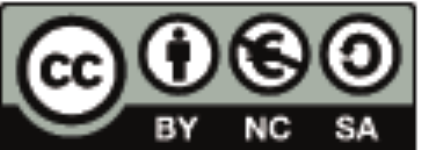}&
  \begin{tabular}{@{}l@{}}
    \textcopyright\,Johannes Ebbing and Peter Lohmann\\
    {\tiny licensed under a \href{http://creativecommons.org/licenses/by-nc-sa/3.0/}{Creative Commons Attribution-NonCommercial-ShareAlike 3.0 Unported License}}
  \end{tabular}
\end{tabular}\\[1.5ex]
This work was partially supported by the NTH Focused Research School for IT Ecosystems.
}

\begin{abstract}
\small
Modal dependence logic (\MDL) was introduced recently by V\"a\"an\"anen. It enhances the basic modal language by an operator $\dep{\cdot}$. For propositional variables $p_{1},\dots,p_{n}$ the atomic formula $\dep[p_{1},\dots,p_{n-1}]{p_{n}}$ intuitively states that the value of $p_{n}$ is determined solely by those of $p_{1},\dots,p_{n-1}$.

We show that model checking for \MDL formulae over Kripke structures is \NP-complete
and further consider fragments of \MDL obtained by restricting the set of allowed propositional and modal connectives. It turns out that several fragments, e.g., the one without modalities or the one without propositional connectives, remain \NP-complete.

We also consider the restriction of \MDL where the length of each single dependence atom is bounded by a number that is fixed for the whole logic.
We show that the model checking problem for this bounded \MDL is still \NP-complete while for some fragments, e.g., the fragment with only $\EX$, the complexity drops to \PTIME.

We additionally extend \MDL by allowing classical disjunction -- introduced by Sevenster -- besides dependence disjunction and show that classical disjunction is always at least as computationally bad as bounded arity dependence atoms and in some cases even worse, e.g., the fragment with nothing but the two disjunctions is \NP-complete.

Furthermore we almost completely classifiy the computational complexity of the model checking problem for all restrictions of propositional and modal operators for both unbounded as well as bounded \MDL with both classical as well as dependence disjunction.

This is the second arXiv version of this paper. It extends the first version by the investigation of the classical disjunction. A shortened variant of the first arXiv version was presented at SOFSEM 2012 \cite{eblo12}.
\end{abstract}

\smallskip
\noindent{\bfseries ACM Subject Classifiers:} \SubjectClassifiers


\section{Introduction}
\newcommand{\fragmenttable}[3]{
\begin{table}[ht]
\begin{center}
\[\begin{array}{cc|ccc|cc|c|p{3cm}}
  \multicolumn{7}{c|}{\textbf{Operators}} & \textbf{Complexity} & \textbf{Reference} \\
  \AX&\EX & \wedge&\vee&\aneg & \dep{}&\nor & &\\
  \hline\hline
#1
\end{array}\]
{\footnotesize $+:$ operator present \quad $-:$ operator absent \quad $*:$ complexity independent of operator}
\vspace*{1ex}
\caption{#2%
}
#3
\end{center}
\end{table}
}

Dependence among values of variables occurs
everywhere in computer science (databases, software engineering, knowledge
representation, AI) but also the social sciences (human history, stock markets,
etc.). In his monograph \cite{va07} in 2007 V\"a\"an\"anen introduced functional dependence
into the language of first-order logic.

Functional dependence of the value of $q$ from the values of $p_1,\ldots,p_n$ means that there exists a determinating function $f$ with
$q = f(p_1,\dots,p_n)$, i.e., the value of $q$ is completely determined by the values of $p_1,\dots,p_n$ alone. We denote this form of dependence (or determination) by the \emph{dependence atom} $\dep[p_1,\dots,p_n]{q}$.
To examine dependence between situations, plays, worlds, events or observations we consider collections of these, so called \emph{teams}. For example, a database can be interpreted as a team. In this case $\dep[p_1,\ldots,p_n]{q}$ means that in every record the value of the attribute $q$ is determined by the values of the attributes $p_1,\dots,p_n$.

In modal logic a team is a set of worlds in a Kripke structure. Here $\dep[p_1,\dots,p_n]{q}$ means that in every world of the team the value of the atomic proposition $q$ is determined by the propositions $p_1,\ldots,p_n$, i.e., there is a fixed Boolean function $f:\{0,1\}^n\to\{0,1\}$ that determines the value of $q$ from the values of $p_1,\dots,p_n$ for all worlds in the team. In first-order logic $\dep[x_1,\dots,x_n]{y}$ means the same for a function $f:A^n\to A$ where $A$ is the universe of a first-order structure. 
\emph{Dependence logic} \cite{va07} is then defined by simply adding dependence atoms to usual first-order logic and \emph{modal dependence logic} (\MDL) \cite{va08,se09} is defined by introducing dependence atoms to modal logic.


Besides the inductive semantics (which we will use here) V\"a\"an\"anen also gave two equivalent game-theoretic semantics for \MDL \cite{va08}. Sevenster showed that for singleton sets of worlds there exists a translation from \MDL to plain modal logic \cite{se09}. Sevenster also showed that the satisfiability problem for \MDL is \NEXPTIME-complete \cite{se09} and Lohmann and Vollmer continued the complexity analysis of the satisfiability problem for \MDL by systematically restricting the set of allowed modal and propositional operators and completely classifying the complexity for all fragments of \MDL definable in this way \cite{lovo10}.

Sevenster \cite{se09} also introduced classical disjunction (which is \emph{classical} in a more set theoretic way of looking at the semantics; cf.~\cite{abva08}) into the language of \MDL. In the following we always think of the version that includes both classical disjunction (here denoted by $\nor$) as well as dependence disjunction when we write \MDL.

\medskip
The method of systematically classifying the complexity of logic related problems by restricting the set of operators allowed in formulae goes back to Lewis who used this method for the satisfiability problem of propositional logic \cite{le79}. Recently it was, for example, used by Hemaspaandra et al.~for the satisfiability problem of modal logic \cite{he05,hescsc10} and by Lohmann and Vollmer for the satisfiability problem of \MDL \cite{lovo10}.
The motivation for this approach is that by systematically examining all fragments of a logic one might find a fragment which allows for efficient algorithms but still has high enough expressivity to be useful in practice.
On the other hand, this systematic approach usually leads to insights into the sources of hardness, i.e., the exact components of the logic that make satisfiability, model checking etc.~hard.

In this paper we transfer the method from satisfiability \cite{lovo10} to model checking and classify the model checking problem for almost all fragments of \MDL definable by restricting the set of allowed modal ($\AX$, $\EX$) and propositional ($\wedge$, $\vee$, $\nor$, $\neg$) operators to an arbitrary subset of all operators.
The model checking problem asks whether a given formula is true in a given team of a given Kripke structure.
For plain modal logic this problem is solvable in \PTIME as shown by Clarke et al.~\cite{clemsi86}. A detailed complexity classification for the model checking problem over fragments of modal logic was shown by Beyersdorff et al.~\cite{bememuscthvo11} (who investigate the temporal logic \CTL which contains plain modal logic as a special case).

In the case of \MDL it turns out that model checking is \NP-complete in general and that this still holds for several seemingly quite weak fragments of \MDL, e.g., the one without modalities or the one where nothing except dependence atoms and $\EX$ is allowed (first and fourth line in \cref{results-unbounded-dep}).
Strangely, this also holds for the case where only the both disjunctions $\sor$ and $\nor$ are allowed and not even dependence atoms occur (third line in \cref{results-unbounded-dep}).

Furthermore it seems natural to not only restrict modal and propositional operators but to also impose restrictions on dependence atoms.
One such restriction is to limit the arity of dependence atoms, i.e., the number $n$ of variables $p_1,\dots,p_n$ by which $q$ has to be determined to satisfy the formula $\dep[p_1,\dots,p_n]{q}$, to a fixed upper bound $k\geq 0$ (the logic is then denoted by \MDLk).
For this restriction model checking remains \NP-complete in general but for the fragment with only the $\EX$ operator allowed this does not hold any more (seventh line in Table~\ref{results-bounded-dep}). In this case either $\wedge$ (fourth line in Table~\ref{results-bounded-dep}) or $\vee$ (sixth line in Table~\ref{results-bounded-dep}) is needed to still get \NP-hardness.

We classify the complexity of the model checking problem for fragments of \MDL with unbounded as well as bounded arity dependence atoms. We are able to determine the tractability of each fragment except the one where formulae are built from atomic propositions and unbounded dependence atoms only by disjunction and negation (sixth line in Table~\ref{results-unbounded-dep}).
In each of the other cases we either show \NP-completeness or show that the model checking problem admits an efficient (polynomial time) solution.

\fragmenttable{
  *&* & +&+&* & +&* & \NP\text{-complete} & \Cref{wedge-vee-np-complete}\\\hline
  +&* & *&+&* & +&* & \NP\text{-complete} & \Cref{box-vee-np-complete}\\\hline 
  *&* & *&+&* & *&+ & \NP\text{-complete} & \Cref{nor vee}\\\hline
  *&+ & *&*&* & +&* & \NP\text{-complete} & \Cref{diamond-np-complete}\\\hline 
  *&+ & +&*&* & *&+ & \NP\text{-complete} & \hbox{\Cref{diamond-wedge-bounded},} \Cref{nor replaces dep}\\\hline
  -&- & -&+&* & +&- & \text{in }\NP & \Cref{mdlmc-in-np}\\\hline
  *&* & -&-&* & -&* & \text{in }\PTIME & \Cref{nor and unary}\\\hline
  *&- & *&-&* & *&* & \text{in }\PTIME & \Cref{box-wedge-in-p}\\\hline 
  *&* & *&*&* & -&- & \text{in }\PTIME & \cite{clemsi86}\\\hline
}{Classification of complexity for fragments of \MDLMC}{\label{results-unbounded-dep}}

\fragmenttable{
  *&* & +&+&* & +&* & \NP\text{-complete} & \Cref{wedge-vee-np-complete}\\\hline
  +&* & *&+&* & +&* & \NP\text{-complete} & \Cref{box-vee-np-complete}\\\hline 
  *&* & *&+&* & *&+ & \NP\text{-complete} & \Cref{nor vee}\\\hline
  *&+ & +&*&* & +&* & \NP\text{-complete} & \Cref{diamond-wedge-bounded}\\\hline 
  *&+ & +&*&* & *&+ & \NP\text{-complete} & \hbox{\Cref{diamond-wedge-bounded},} \Cref{nor replaces dep}\\\hline
  *&+ & *&+&* & +&* & \NP\text{-complete} & \Cref{diamond-vee-bounded}\\\hline 
  *&* & -&-&* & *&* & \text{in }\PTIME & 
  \Cref{nor and unary} \\\hline 
  *&- & *&-&* & *&* & \text{in }\PTIME & \Cref{box-wedge-in-p}\\\hline 
  -&- & -&*&* & *&- & \text{in }\PTIME & \Cref{vee-bounded-in-p} \\\hline
  *&* & *&*&* & -&- & \text{in }\PTIME & \cite{clemsi86}\\\hline
}{Classification of complexity for fragments of \MDLMCk with $k\geq 1$}{\label{results-bounded-dep}}

In Table~\ref{results-unbounded-dep} we list our complexity results 
for the cases with unbounded arity dependence atoms and in Table~\ref{results-bounded-dep} for the cases with an a priori bound on the arity.
In these tables a \anfz{+} means that the operator is allowed, a \anfz{-} means that the operator is forbidden and a \anfz{*} means that the operator does not affect the complexity of the problem.


\section{Modal dependence logic}
We will briefly present the syntax and semantics of \MDL. For a more in-depth introduction we refer to V\"a\"an\"anen's definition of \MDL \cite{va08} and Sevenster's model-theoretic and complexity analysis \cite{se09} which also contains a self-contained introduction to \MDL.

\begin{definition}\label{defMDLsyn}(Syntax of \MDL)\\
Let $AP$ be an arbitrary set of atomic propositions and $p_1,\ldots,p_n,q \in AP$. Then \MDL is the set of all formulae built from the following rules:
\[\begin{array}{rcl}
\varphi & \ddfn & \true \;\mid\; \false \;\mid\; q \;\mid\; \neg q \;\mid\; \varphi\vee \varphi \;\mid\; \varphi\nor\varphi \;\mid\; \varphi\wedge\varphi \;\mid\; \AX\varphi\; \;\mid\; \EX\varphi\; \;\mid\\
        &       & \dep[p_1,\ldots,p_n]{q} \;\mid\;\neg\dep[p_1,\ldots,p_n]{q}.
\end{array}\]
Note that negation 
is only atomic, i.e., it is only defined for atomic propositions and dependence atoms.
\end{definition}

We sometimes write $\AX^k$ (resp.~$\EX^k$) for $\underbrace{\AX\AX\dots\AX}_{k \text{ times}}$ (resp.~$\underbrace{\EX\EX\dots\EX}_{k \text{ times}}$).
For a dependence atom $\dep{p_1,\dots,p_n,q}$ we define its \emph{arity} as $n$, i.e., the arity of a dependence atom is the arity of the determinating function whose existence it asserts.

In Section~\ref{sec:bounded-fragments} we will investigate the model checking problem for the following logic.

\begin{definition}\label{defMDLk}(\MDLk)\\
\MDLk is the subset of \MDL that contains all formulae which do not contain any dependence atoms whose arity is greater than $k$.
\end{definition}

We will classify \MDL for all fragments defined by sets of operators.

\begin{definition}\label{defOpFrag}(\MDL[M])\\
Let $M\subseteq \{\AX,\EX,\wedge,\vee,\nor,\aneg,\true,\false,\dep{}\}$. By \MDL[M] (resp.~\MDLk[M]) we denote the subset of \MDL (resp.~\MDLk) built from atomic propositions using only operators from $M$. We sometimes write $\MDL[op1,op2,\dots]$ instead of $\MDL[\{op1,op2,\dots\}]$.
\end{definition}

\MDL formulae are interpreted over Kripke structures.

\begin{definition}\label{defKripke}(Kripke structure)\\
An \emph{$AP$-Kripke structure} is a tuple $W=(S,R,\pi)$ where $S$ is an arbitrary non-empty set of \emph{worlds}, $R\subseteq S\times S$ is the \emph{accessibility relation} and $\pi:S \to \powerset{AP}$ is the \emph{labeling function}.
\end{definition}

\begin{definition}\label{defMDLsem}(Semantics of \MDL)\\
In contrast to common modal logics, truth of a \MDL formula is not defined with respect to a single world of a Kripke structure but with respect to a set (or \emph{team}) of worlds.
Let $AP$ be a set of atomic propositions and $p, p_1, \dots, p_n \in AP$. The \emph{truth} of a formula $\varphi\in\MDL$ in a team $T\subseteq S$ of an $AP$-Kripke structure $W=(S,R,\pi)$ is denoted by $W,T\models \varphi$ and is defined as follows:
\[\begin{array}{lcl@{}c@{\,}p{7.6cm}}
W,T&\models&\true&\quad\quad&always holds\\
W,T&\models&\false&\quad\text{iff}\quad&$T=\emptyset$\\
W,T&\models&p&\quad\text{iff}\quad&$p\in\pi(s)$ for all $s\in T$\\
W,T&\models&\neg p&\quad\text{iff}\quad&$p\notin\pi(s)$ for all $s\in T$\\
W,T&\models&\dep[p_1,\dots,p_{n-1}]{p_n}&\quad\text{iff}\quad&for all $s_1,s_2\in T$ it holds that\newline
  $\pi(s_1)\cap\{p_1,\dots,p_{n-1}\}\neq\pi(s_2)\cap\{p_1,\dots,p_{n-1}\}$\newline
  or $\pi(s_1)\cap\{p_n\} = \pi(s_2)\cap\{p_n\}$\\
W,T&\models&\neg\dep[p_1,\dots,p_{n-1}]{p_n}&\quad\text{iff}\quad&$T=\emptyset$\\
W,T&\models&\varphi\wedge\psi&\quad\text{iff}\quad&$W,T\models \varphi$ and $W,T\models \psi$\\
W,T&\models&\varphi\nor\psi&\quad\text{iff}\quad&$W,T\models \varphi$ or $W,T\models \psi$\\
W,T&\models&\varphi\vee\psi&\quad\text{iff}\quad&there are sets $T_1,T_2$ with $T=T_1\cup T_2$,\ \,$W,T_1\models\varphi$ and $W,T_2\models \psi$\\
W,T&\models&\AX \varphi&\quad\text{iff}\quad&$W,\{s'\mid \exists s\in T$ with $(s,s')\in R\}\models \varphi$\\
W,T&\models&\EX \varphi&\quad\text{iff}\quad&there is a set $T'\subseteq S$ such that $W,T'\models \varphi$ and for all $s\in T$ there is a $s'\in T'$ with $(s,s')\in R$
\end{array}\]

%
%
\end{definition}
Note that this semantics is a conservative extension of plain modal logic semantics, \ie it coincides with the latter for formulae which do neither contain dependence atoms nor classical disjunction. Rationales for this semantics -- especially for the case of the negative dependence atom -- were given by V\"a\"an\"anen \cite[p.~24]{va07}.

In the remaining \lcnamecrefs{sec:unbounded-fragments} we will classify the complexity of the model checking problem for fragments of \MDL and \MDLk.

\begin{definition}
\label{defMDLMC}(\MDLMC)\\
Let $M\subseteq \{\AX,\EX,\wedge,\vee,\nor,\aneg,\true,\false,\dep{}\}$. Then the model checking problem for \MDL[M] (resp.~\MDLk[M]) over Kripke structures is defined as the canonical decision problem of the set
\[\begin{aligned}
\begin{array}{@{}l@{}}
    \MDLMC[M]\\
    \text{(resp.~$\MDLMCk[M]$)}
    \end{array}
  &  \dfn \set[7cm]{\enc{W,T,\varphi}}{$W=(S,R,\pi)$ a Kripke structure, $T\subseteq S$, $\varphi\in\MDL[M]$ and $W,T\models \varphi$}.\\
\end{aligned}\]
We write \MDLMC for $\MDLMC[\{\AX, \EX,\wedge, \vee,\aneg,\dep{},\nor,\true,\false\}]$.

\end{definition}

Before we begin with the classification we state a lemma showing that it does not matter whether we include $\top$, $\bot$ or $\neg$ in a sublogic \MDL[M] of \MDL since this does not affect the complexity of \MDLMC[M].
\begin{lemma}\label{neg-dont-matter}
Let $M$ be an arbitrary set of \MDL operators, \ie $M\subseteq\{\allMDL,\allowbreak\bot,\top\}$. Then we have that 
\[\MDLMC[M]\ \equivpm\ \MDLMC[M\setminus \{\true,\false,\neg\}].\]
\end{lemma}
\begin{proof}
It suffices to show $\leqpm$.
So let $K=(S,R,\pi)$ a Kripke structure, $T\subseteq S$, $\varphi \in \MDL[M]$ and the variables of $\varphi$ among $p_1,\dots,p_n$.
Let $p_1',\dots,p_n',t,f$ be fresh propositional variables.
Then $K,T \models \varphi$ iff $K',T\models \varphi'$, where
$K' \dfn (S,R,\pi')$ 
with $\pi'$ defined by
\[\begin{array}{rcl}
\pi'(s)\cap \{t,f\}            &\dfn& \{t\},\\
\pi'(s)\cap \{p_i,p_i'\}      &\dfn& \left\{
                         \begin{array}{ll}
                           \{p_i\}          & \text{if }p_i\in\pi(s),\\
                           \{p_i'\}       & \text{if }p_i\notin\pi(s),
                         \end{array}
                       \right.
\end{array}\]
for all $i\in\{1,\dots,n\}$ and $s\in S$, and $\varphi \in \MDL[M\setminus \{\true,\false,\neg\}]$ defined by
\[\varphi' \ \dfn\  \varphi(p_1'/\neg p_1)(p_2'/\neg p_2)\dots(p_n'/\neg p_n)(t / \true)(f / \false).\]
\end{proof}

\section{Unbounded arity fragments}\label{sec:unbounded-fragments}

First we will show that the most general of our problems is in \NP and therefore all model checking problems investigated later are as well.

\begin{proposition}\label{mdlmc-in-np}
Let $M$ be an arbitrary set of \MDL operators. Then \MDLMC[M] is in \NP. And hence also $\MDLMCk[M]$ is in  $\NP$ for every $k\geq0$.
\end{proposition}
\begin{proof}
The following non-deterministic top-down algorithm 
checks the truth of the formula $\varphi$ on the Kripke structure $W$ in the evaluation set $T$ in polynomial time.

\Needspace{6\baselineskip}
\begin{lstlisting}[caption={\lstinline!check($W=(S,R,\pi)$, $\varphi$, $T$)!}, label={algo:check}, escapechar={§}]
$%bool check($W$, $T$, $\varphi$)
%
$case $\varphi$
when $\varphi=p$
  foreach $s \in T$
    if not $p \in \pi(s)$ then
      return false
  return true
§\Needspace{5\baselineskip}§
when $\varphi=\neg p$
  foreach $s \in T$
    if $p \in \pi(s)$ then
      return false
  return true
§\Needspace{7\baselineskip}§
when $\varphi = \,\dep{p_1,\ldots,p_n}$
  foreach $(s,s')\in T\times T$
    if $\pi(s)\cap\{p_1,\dots,p_{n-1}\}\,=\,\pi(s')\cap\{p_1,\dots,p_{n-1}\}$ then
          // i.e., $s$ and $s'$ agree on the values of the propositions $p_1,\dots, p_{n-1}$
      if ($q\in\pi(s)$ and not $q\in \pi(s')$) or (not $q\in\pi(s)$ and $q\in \pi(s')$) then
        return false
  return true
§\Needspace{4\baselineskip}§
when $\varphi = \, \neg\dep{p_1,\ldots,p_n}$
  if $T=\emptyset$
    return true
  return false
§\Needspace{6\baselineskip}§
when $\varphi = \psi\vee \theta$
  guess two sets of states $A,\,B \subseteq S$
  if not $A\cup B=T$ then
    return false
  return (check($W,A,\psi$) and check($W,B,\theta$))
§\Needspace{2\baselineskip}§
when $\varphi = \psi\nor \theta$
  return (check($W,T,\psi$) or check($W,T,\theta$))
§\Needspace{2\baselineskip}§
when $\varphi= \psi\wedge \theta$
  return (check($W,T,\psi$) and check($W,T,\theta$))
§\Needspace{8\baselineskip}§
when $\varphi= \AX \psi$
  $T':=\emptyset$
  foreach $s' \in S$
    foreach $s \in T$
      if $(s,s') \in R$ then
        $T':= T' \cup \{s'\}$
            // $T'$ is the set of all successors of all states in $T$
  return check($W,T',\psi$)
§\Needspace{7\baselineskip}§
when $\varphi= \EX \psi$
  guess set of states $T'\subseteq S$
    foreach $s\in T$
      if there is no $s'\in T'$ with $(s,s')\in R$ then
        return false
            // $T'$ contains at least one successor of every state in T
  return check($W,T',\psi$)
\end{lstlisting}
\end{proof}

Now we will see that the model checking problem is \NP-hard and that this still holds without modalities.

\begin{theorem}\label{wedge-vee-np-complete}
Let $M\supseteq\{\wedge, \vee, \dep{}\}$. Then \MDLMC[M] is \NP-complete. Furthermore, \MDLMCk[M] is \NP-complete for every $k\geq 0$.
\end{theorem}
\begin{proof}
Membership in \NP follows from Proposition \ref{mdlmc-in-np}. For the hardness proof we reduce from \ThreeSAT.

For this purpose let $\varphi = C_{1}\wedge\ldots\wedge C_{m}$
be an arbitrary $\CNF$ formula
with variables $x_{1},\ldots,x_{n}$. Let $W$ be the Kripke structure $\left(S,R,\pi\right)$ over the atomic propositions $r_1,\dots,r_n,p_1,\dots,p_n$ defined by
\[\begin{array}{lcl}
S	&	\dfn & \{ s_{1}, \ldots, s_{m} \},\\
R & \dfn & \emptyset,\\
\pi(s_{i}) \cap \{r_j,p_j\} & \dfn & \begin{cases}
\{r_j,p_j\} &\mbox{ iff $x_j$ occurs in $C_{i}$ positively,}\\
\{r_j\} &\mbox{ iff $x_j$ occurs in $C_{i}$ negatively,}\\
\emptyset &\mbox{ iff $x_j$ does not occur in $C_{i}$.}\\
\end{cases}
\end{array}\]

Let $\psi$ be the $\MDL[\wedge,\vee,\dep{}_0]$ formula
\[\bigvee_{j=1}^{n}\ r_j\wedge \dep{p_j}\]
and let $T\dfn\left\{s_{1},\ldots,s_{m}\right\}$ the evaluation set.

We will show that $\varphi \in \ThreeSAT$ iff $W, T \models \psi$. Then it follows that $\ThreeSAT\leqpm \MDLMCpara[M]{0}$ and therefore $\MDLMCpara[M]{0}$ is \NP-hard.

Now assume $\varphi \in \ThreeSAT$ and $\theta$ an interpretation with $\theta \models \varphi$. 
From the valuations $\theta(x_j)$ of all $x_j$ we construct subteams $T_1,\ldots, T_n$ such that for all $j \in \{1,\ldots, n\}$ it holds that $W, T_j \models \gamma_j$ with $\gamma_j \dfn r_j \wedge \dep{p_j}$. The $T_j$ are constructed as follows
\begin{align*}
&T_j \dfn \begin{cases}
\{s_i \in S \mid \pi(s_i)\cap \{r_j, p_j\} = \{r_j, p_j\}\} &\text{ iff } \theta(x_j) = 1\\
\{s_i \in S \mid \pi(s_i)\cap \{r_j, p_j\} = \{r_j\}\} &\text{ iff } \theta(x_j) = 0
\end{cases}\\
&\text{i.e., } T_j \text{ is the team consisting of exactly the states}\\
&\text{corresponding to clauses satisfied by } \theta(x_j).
\end{align*}
Since every clause in $\varphi$ is satisfied by some valuation $\theta(x_j) = 1$ or $\theta(x_j) = 0$ we have that $T_1 \cup \ldots \cup T_n = T$ such that $W, T \models \varphi$.

On the other hand, assume that $W, T \models \psi$, therefore we have $T = T_1 \cup T_2 \cup \ldots \cup T_n$ such that for all $j \in \{1,\ldots,n\}$ it holds that $T_j \models \gamma_j$. Therefore $\pi(s_i)\cap\{p_j\}$ is constant for all elements $s_i \in T_j$. From this we can construct a valid interpretation $\theta$ for $\varphi$. 

For all $j$ let $I_j \dfn \{i \mid s_i \in T_j\}$. For every $j\in \{1,\ldots,n\}$ we consider $T_j$. If for every element $s_i \in T_j$ it holds that $\pi(s_i)\cap \{p_j\} = \{p_j\}$ then we have for all $i \in I_j$ that $x_j$ is a literal in $C_i$. In order to satisfy those $C_i$ we set $\theta(x_j)=1$. If for every element $s_i \in T_j$ it holds that $\pi(s_i) \cap \{p_j\} = \emptyset$ then we have for every $i \in I_j$ that $\neg {x_j}$ is a literal in $C_i$. In order to satisfy those $C_i$ we set $\theta(x_j) = 0$.\\
Since for every $s_i \in T$ there is a $j$ with $s_i \in T_j$ we have an evaluation $\theta$ that satisfies every clause in $\varphi$. Therefore we have $\theta \models \varphi$.
\end{proof}

Instead of not having modalities at all we can also allow nothing but the $\EX$ modality, i.e., we disallow propositional connectives and the $\AX$ modality, and model checking is \NP-complete as well.

\begin{theorem}
\label{diamond-np-complete}
Let $M\supseteq \{\EX,\dep{}\}$. Then \MDLMC[M] is \NP-complete.
\end{theorem}
\begin{proof}
Membership in \NP follows from Proposition \ref{mdlmc-in-np} again.

For hardness we again reduce from $\ThreeSAT$.
Let $\varphi = \bigwedge_{i=1}^{m}C_i$ be an arbitrary \CNF formula built from the variables $x_1,\ldots,x_n$. Let $W$ be the Kripke structure $(S,R,\pi)$, over the atomic propositions $p_1,\dots,p_n,q$, shown in Figure~\ref{figure:diamond-dep1} 
and formally defined by
\[\begin{array}{lcl}
S	&	\dfn & \{c_1,\ldots,c_m,s_1^1,\ldots,s_n^1, s_1^0 ,\ldots, s_n^0  \},\\
R \cap\{(c_i,s_j^1),(c_i,s_j^0)\} & \dfn & \begin{cases}
\{(c_i, s_j^1)\}  & \text{ iff $x_j$ occurs in $C_{i}$ positively,}\\
\{(c_i, s_j^0 )\} & \text{ iff $x_j$ occurs in $C_{i}$ negatively,}\\
\emptyset         & \text{ iff $x_j$ does not occur in $C_{i}$,}\\
\end{cases}\\
\pi(c_i) & \dfn & \emptyset,\\
\pi(s_j^1)  &\dfn & \{p_j,q\},\\
\pi(s_j^0 ) &\dfn & \{p_j\}.
\end{array}\]

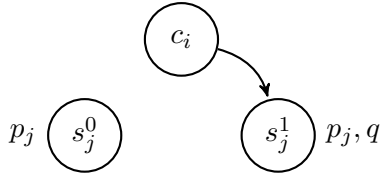
\begin{figure}[ht]
\begin{center}
\begin{tikzpicture}[->,>=stealth',shorten >=1pt,auto,node distance=1.8cm,
                    thick]
                    
 	\node[state] 				 (C1)                      {$c_i$};
  \node[state]    					     (x1_) [below left of=C1, label=left:{$p_j$}]  {$s_j^0$};
  \node[state]        					 (x1)  [below right of=C1, label=right:{$p_j,q$}] {$s_j^1$};

  \path (C1) edge [bend left] node {}(x1);
\end{tikzpicture}
\caption{Kripke structure part corresponding to the \CNF fragment $\dots \wedge C_i \wedge \dots$ with $C_i = x_j \vee \dots$.}
\label{figure:diamond-dep1}
\end{center}
\end{figure}

%
%
%

Let $\psi$ be the $\MDL[\EX,\dep{}]$ formula
\[\EX  \dep[p_1,\ldots, p_n]{q}\]
and let $T\dfn\{c_1,\ldots,c_m\}$.

We will show that $\varphi \in \ThreeSAT$ iff $W, T \models \psi$. Hence, $\ThreeSAT\leqpm \MDLMC[M]$ and $\MDLMC[M]$ is \NP-hard.

First suppose we have an interpretation $\theta$ that satisfies $\varphi$. From the valuations of $\theta$ we will construct a successor team $T'$ of $T$, i.e., for all $s\in T$ there is an $s'\in T'$ s.t. $(s,s')\in R$ with $W, T' \models \dep{p_1,\ldots, p_n,q}$. $T'$ is defined by:
\[
T' \dfn \{s_j^z \mid \theta(x_j) = z, j \in \{1,\ldots,n\}\}
\]
Since $\theta$ satisfies every clause $C_i$ of $\varphi$ we have that for every $C_i$ there is an $x_j$ with 
\[
\theta(x_j) = \begin{cases}
1, &\text{iff } x_j \in C_i\\
0, &\text{iff } \neg x_j \in C_i.
\end{cases}
\]
It follows that for every $s \in T$ there is an $s' \in T'$ such that $(s,s')\in R$.

By construction of $T'$ it is not possible to have both $s_j^0$ and $s_j^1$ in $T'$. Hence for all elements $s_j^0, s_{j'}^{1} \in T'$ it follows that $j\neq j'$ and therefore $\pi(s_j^0)\cap \{p_1,\ldots,p_n\} \neq \pi(s_{j'}^{1})\cap\{p_1,\ldots,p_n\}$. Thus $W, T' \models \dep{p_1,\ldots, p_n,q}$.


On the other hand assume $W,T\models \psi$. Then there is a successor set $T'$ of $T$ s.t.~for every $s \in T$ there is an $s' \in T'$ with $(s,s')\in R$ and $T'\models \dep{p_1,\ldots, p_n,q}$. We construct $\theta$ as follows:
\[
\theta(x_j) \dfn \begin{cases}
1, &\text{iff } s_j^1\in T'\\
0, &\text{iff } s_j^0 \in T'\\
0, &\text{iff } s_j^0,s_j^1 \notin T'.
\end{cases}
\]
Note that in the latter case it does not matter if 0 or 1 is chosen.

Since $W,T' \models \dep{p_1,\ldots, p_n, q}$ and for every $j$ it holds that $W,\{s_j^0,s_j^1\}\not\models \dep{p_1,\ldots,p_n,\allowbreak q}$ we have that for every $j$ at most one of $s_j^0$ or $s_j^1$ is in $T'$. It follows that $\theta$ is well-defined.

Since for every $c_i\in T$ there is an $s_j^z \in T'$ s.t. $(c_i,s_j^z) \in R$ with $\theta(x_j) = z$, we have by contruction of $W$ that $\theta$ satisfies every clause $C_i$ of $\varphi$. From this follows $\varphi \in \ThreeSAT$.
\end{proof}

If we disallow $\EX$ but allow $\AX$ instead we have to also allow $\vee$ to get \NP-hardness.

\begin{theorem}
\label{box-vee-np-complete}
Let $M\supseteq \{\AX,\vee,\dep{}\}$. Then \MDLMC[M] is \NP-complete. Also, \MDLMCk[M] is \NP-complete for every $k\geq 0$.
\end{theorem}
\begin{proof}
Membership in \NP follows from Proposition \ref{mdlmc-in-np} again.
To prove hardness, we will once again reduce \ThreeSAT to this problem.

Let $\varphi = \bigwedge_{i=1}^{m}C_i$ be an arbitrary $\CNF$ formula over the variables $x_1,\ldots,x_n$.
Let $W$ be the structure $(S,R,\pi)$, over the atomic propositions $p_1,\dots,p_n$, shown in Figures~\ref{chain1} to \ref{chain6} and formally defined as follows:
\[\begin{array}{lcl}
S & \dfn & \big\{s_i | i \in \{1,\ldots,m\}\big\}\\
  &      & \cup \ \big\{r_k^j | k\in \{1,\ldots,m\}, j \in \{1,\ldots,n\}\big\} \\
  &      & \cup \ \big\{\overline r_k^j | k\in \{1,\ldots,m\}, j \in \{1,\ldots,n\}\big\}\\[2ex]

\multicolumn{3}{l}{R\, \cap\bigcup\limits_{j\in\{1,\dots,n\}}\,\{(s_i,r_i^j),(s_i,\overline r_i^j)\}\ \dfn}\\
\multicolumn{3}{l}{\qquad\left\{\begin{array}{lll}
\{(s_i,r_i^1)\}                                         & \text{iff $x_1$ occurs in $C_i$ (positively or negatively)} & \text{\emph{(Fig.~\ref{chain1})}}\\
\{(s_i,r_i^1),(s_i, \overline r_i^1 )\}                 & \text{iff $x_1$ does not occur in $C_i$} & \text{\emph{(Fig.~\ref{chain2})}}\\
\end{array}\right.}\\[4ex]

\multicolumn{3}{l}{R \,\cap \bigcup\limits_{k\in\{1,\dots,n\}}\,\{(r_i^j,r_i^{k}),(r_i^j,\overline r_i^{k}),(\overline r_i^j,r_i^{k}),(\overline r_i^j,\overline r_i^{k})\}\ \dfn}\\
\multicolumn{3}{l}{\qquad\left\{\begin{array}{lll}
\{(r_i^j,r_i^{j+1})\}                                   & \parbox[t]{15em}{iff $x_{j}$ and $x_{j+1}$ both occur in $C_i$} & \text{\emph{(Fig.~\ref{chain3})}}\\
\{(r_i^j,r_i^{j+1}),(r_i^j, \overline r_i^{j+1} )\}     & \parbox[t]{15em}{iff $x_{j}$ occurs in $C_i$ but $x_{j+1}$ does not occur in $C_i$} & \text{\emph{(Fig.~\ref{chain4})}}\\[3ex]
\{(r_i^j,r_i^{j+1}),(\overline r_i^j ,r_i^{j+1})\}      & \parbox[t]{15em}{iff $x_j$ does not occur in $C_i$ but $x_{j+1}$ does occur in $C_i$} & \text{\emph{(Fig.~\ref{chain5})}}\\[3ex]
\{(r_i^j,r_i^{j+1}),(\overline r_i^j , \overline r_i^{j+1} )\} & \parbox[t]{15em}{iff neither $x_j$ nor $x_{j+1}$ occur in $C_i$} & \text{\emph{(Fig.~\ref{chain6})}}
\end{array}\right.}\\[11ex]

\pi(s_i) & \dfn & \emptyset\\
\pi(r_i^j) & \dfn & \begin{cases}\{p_j\}   &\text{iff $x_j$ occurs in $C_i$ positively or not at all}\\
                                 \emptyset &\text{iff $x_j$ occurs in $C_i$ negatively}\end{cases}\\
\pi(\overline r_i^j ) & \dfn & \emptyset
\end{array}\]

\begin{figure}
\begin{minipage}[b]{0.48\textwidth}
\centering
\begin{tikzpicture}[->,>=stealth',shorten >=1pt,auto,node distance=2.8cm,
                    thick]
	\node[state]						 				 	 (C1)                   		{$s_i$};
	\node[state]											 (s1x1)  [below of=C1]	 		{$r_i^1$};
	
	\path (C1) 				edge       node {}(s1x1);
	
\end{tikzpicture}
\caption{$x_1$ occurs in $C_i$.}
\label{chain1}
\end{minipage}
\hspace{0.02\textwidth}
\begin{minipage}[b]{0.48\textwidth}
\centering
\begin{tikzpicture}[->,>=stealth',shorten >=1pt,auto,node distance=2.8cm,
                    thick]
	\node[state]						 				 	 (C1)                   				{$s_i$};
	\node[state]											 (s1x1)  [below right of=C1] 		{$r_i^1$};
	\node[state]											 (s1x1_) [below left of=C1]	  	{$\overline r_i^1 $};
	
	\path (C1) 				edge       node {}(s1x1_)
						 				edge       node {}(s1x1);
\end{tikzpicture}
\caption{$x_1$ does not occur in $C_i$.}
\label{chain2}
\end{minipage}
\end{figure}

\begin{figure}
\begin{minipage}[b]{0.48\textwidth}
\begin{center}
\begin{tikzpicture}[->,>=stealth',shorten >=1pt,auto,node distance=2.8cm,
                    thick]
	\node[state]											 (s1x2)  										 			{$r_i^{j+1}$};
	\node[state]						 				 	 (s1x1)  [above of=s1x2]    			{$r_i^j$};
	
	\path (s1x1) 				edge       node {}(s1x2);
\end{tikzpicture}
\caption{$x_j$ and $x_{j+1}$ occur in $C_i$.}
\label{chain3}
\end{center}
\end{minipage}
\hspace{0.02\textwidth}
\begin{minipage}[b]{0.48\textwidth}
\begin{center}
\begin{tikzpicture}[->,>=stealth',shorten >=1pt,auto,node distance=2.8cm,
                    thick]
	\node[state]						 				 	 (s1x1)      											{$r_i^j$};
	\node[state]						 				 	 (s1x2_) [below left of=s1x1]   	{$\overline r_i^{j+1 }$};
	\node[state]											 (s1x2)  [below right of=s1x1]		{$r_i^{j+1}$};
	
	\path (s1x1) 				edge       node {}(s1x2);
  \path (s1x1)				edge       node {}(s1x2_);
\end{tikzpicture}
\caption{$x_j$ occurs in $C_i$ but $x_{j+1}$ does not occur in $C_i$.}
\label{chain4}
\end{center}
\end{minipage}
\end{figure}

\begin{figure}
\begin{minipage}[b]{0.48\textwidth}
\begin{center}
\begin{tikzpicture}[->,>=stealth',shorten >=1pt,auto,node distance=2.8cm,
                    thick]
	\node[state]											 (s1x2)  										 			{$r_i^{j+1}$};
	\node[state]						 				 	 (s1x1)  [above right of=s1x2]    {$r_i^j$};
	\node[state]						 				 	 (s1x1_) [above left of=s1x2]   	{$\overline r_i^j $};

	\path (s1x1) 				edge       node {}(s1x2);
  \path (s1x1_)				edge       node {}(s1x2);
\end{tikzpicture}
\caption{$x_j$ does not occur in $C_i$ but $x_{j+1}$ does occur in $C_i$.}
\label{chain5}
\end{center}
\end{minipage}
\hspace{0.02\textwidth}
\begin{minipage}[b]{0.48\textwidth}
\begin{center}
\begin{tikzpicture}[->,>=stealth',shorten >=1pt,auto,node distance=2.8cm,
                    thick]
	\node[state]						 				 	 (s1x1)  											    {$r_i^j$};
	\node[state]						 				 	 (s1x1_) [left of=s1x1]  			 		{$\overline r_i^j $};
	\node[state]											 (s1x2)  [below of=s1x1] 		 			{$r_i^{j+1}$};	
	\node[state]											 (s1x2_) [below of=s1x1_]		 			{$\overline r_i^{j+1} $};	
	
	\path (s1x1) 				edge       node {}(s1x2);
  \path (s1x1_)				edge       node {}(s1x2_);
\end{tikzpicture}
\caption{$x_j$ and $x_{j+1}$ do not occur in $C_i$.}
\label{chain6}
\end{center}
\end{minipage}
\end{figure}

Let $\psi$ be the $\MDL[\AX,\vee,\dep{}]$ formula
\[\bigvee_{j=1}^n\ \AX^{j} \dep{p_j}\]
and let $T \dfn \{s_1,\ldots,s_m\}$.

Then, as we will show, $\varphi \in \ThreeSAT$ iff $W,T \models \psi$ and therefore $\MDLMCpara[\AX,\vee,\dep{}]{0}$ is \NP-complete.
Intuitively, the direction from left to right holds because the disjunction splits the team $\{s_1,\dots,s_m\}$ of all starting points of chains of length $n$ into $n$ subsets (one for each variable) in the following way: $s_i$ is in the subset that belongs to $x_j$ iff $x_j$ satisfies the clause $C_i$ under the variable valuation that satisfies $\varphi$. Then the team that belongs to $x_j$ collectively satisfies the disjunct $\AX^j \dep{p_j}$ of $\psi$. For the reverse direction the $\overline r_i^j$ states are needed to ensure that a state $s_i$ can only satisfy a disjunct $\AX^j \dep{p_j}$ if there is a variable $x_j$ that occurs in clause $C_i$ (positively or negatively) and satisfies $C_i$.

More precisely, assume $\theta$ is a satisfying interpretation for $\varphi$. From $\theta$ we construct subteams $T_1,\ldots, T_n$ with $T_1 \cup \ldots \cup T_n = T$ s.t.~for all $j$ it holds that $T_j \models \AX^j \dep{p_j}$. $T_j$ is defined by
\[
T_j \dfn \begin{cases}
\{s_i  \mid \{s_i\} \models \AX^j p_j\} &\text{ iff } \theta(x_j) = 1\\
\{s_i  \mid \{s_i\} \models \AX^j \neg p_j\} &\text{ iff } \theta(x_j) = 0
\end{cases}
\] for all $j\in\{1,\dots,n\}$.
Obviously, for all $j$ it holds that $T_j \models \AX^j \dep{p_j}$.
Now we will show that for all $i\in\{1,\dots,m\}$ there is a $j\in\{1,\dots,n\}$ such that $s_i \in T_j$.
For this purpose let $i\in\{1,\dots,m\}$ and suppose $C_i$ is satisfied by $\theta(x_j)=1$ for a $j\in\{1,\dots,n\}$. Then, by definition of $W$, $\pi(r_i^j)=p_j$, hence $\{s_i\}\models \AX^j p_j$ and therefore $s_i \in T_j$. If, on the other hand, $C_i$ is satisfied by $\theta(x_j)=0$ then we have that $\pi(r_i^j) = \emptyset$, hence $\{s_i\}\models \AX^j \neg p_j$ and again it follows that $s_i \in T_j$.
Altogether we have that for all $i$ there is a $j$ such that $s_i \in T_j$. It follows that $T_1 \cup \ldots \cup T_n = T$ and therefore $W,T \models \psi$.

On the other hand assume $W,T \models \psi$. Therefore we have $T = T_1 \cup \ldots \cup T_n$ with $T_j \models \AX^j \dep{p_j}$ for all $j\in\{1,\dots,n\}$. We define a valuation $\theta$ by
\[
\theta(x_j) \dfn \begin{cases}
1 &\text{iff } T_j \models \AX^j p_j\\
0 &\text{iff } T_j \models \AX^j \neg p_j.
\end{cases}
\]
Since every $s_i$ is contained in a $T_j$ we know that for all $i\in\{1,\dots,m\}$ there is a $j\in\{1,\dots,n\}$ with $\{s_i\} \models \AX^j \dep{p_j}$. From this it follows that $x_j$ occurs in $C_i$ (positively or negatively) since otherwise, by definition of $W$, both $r_i^j$ and $\overline r_i^j$ would be reachable from $s_i$.

It also holds that $\{s_i\} \models \AX^j p_j$ or $\{s_i\} \models \AX^j \neg p_j$. In the former case we have that $\pi(r_i^j) = p_j$, hence, by definition of $W$, $x_j$ is a literal in $C_i$. By construction of $\theta$ it follows that $C_i$ is satisfied. In the latter case it holds that $\overline x_j$ is a literal in $C_i$. Again, by construction of $\theta$ it follows that $C_i$ is satisfied. Hence, $\varphi \in \ThreeSAT$.
\end{proof}

The following example demonstrates the construction from the previous proof.
\begin{example}\label{box-vee-np-complete-example}
Let $\varphi$ be the \CNF formula
\[\underbrace{(\neg x_1 \vee x_2 \vee x_3)}_{C_1}\,\wedge\,\underbrace{(x_2\vee \neg x_3 \vee x_4)}_{C_2}\,\wedge\,\underbrace{(x_1 \vee \neg x_2)}_{C_3}.\]
The corresponding Kripke structure $W$ shown in Figure~\ref{fig:wedge-vee-np-complete-example} has \emph{levels} 0 to 4 where the $j$th level (corresponding to the variable $x_j$ in the formula $\varphi$) is the set of nodes reachable via exactly $j$ transitions from the set of nodes $s_1$, $s_2$ and $s_3$ (corresponding to the clauses of $\varphi$). In this example all non connected states (which do not play any role at all) are not shown.

The \MDL formula corresponding to $\varphi$ is
\[\psi\ =\ \underbrace{\AX \dep{p_1}}_{\gamma_1} \vee \underbrace{\AX\AX \dep{p_2}}_{\gamma_2} \vee \underbrace{\AX\AX\AX \dep{p_3}}_{\gamma_3} \vee \underbrace{\AX\AX\AX\AX \dep{p_4}}_{\gamma_4}.\]

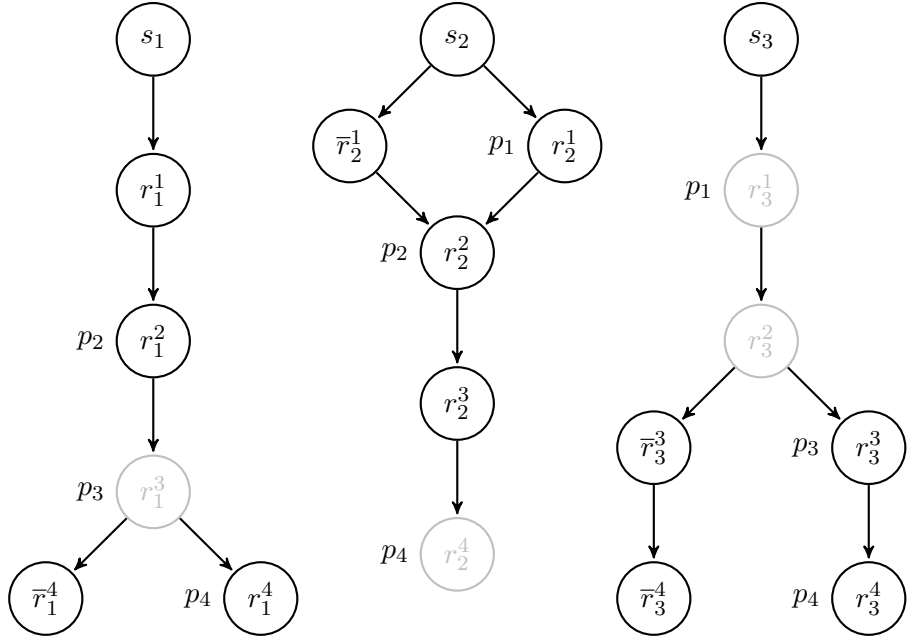
\begin{figure}
\begin{center}
\begin{tikzpicture}[->,>=stealth',shorten >=1pt,auto,node distance=2.0cm,
                    thick]
        \node[state]                                         (C1)                                                                                                                           {$s_1$};
        \node[color=white, state]                                (C1_)   [right of=C1]                                                                  {$s_1$};
  \node[state]               (C2)    [right of=C1_]                                                                                                                 {$s_2$};
  \node[color=white, state]                              (C2_)   [right of=C2]                                                                  {$s_1$};
  \node[state]                 (C3)    [right of=C2_]                                                                                                               {$s_3$};
  \node[state]                                                                                   (s1x1)  [below of=C1, label=left:{}]              {$r_1^1$};
  \node[state]                                                           (s2x1_) [below left of=C2]                                                                                             {$\overline r_2^1 $};
  \node[state]                                                                                   (s2x1)  [below right of=C2, label=left:$p_1$]                  {$r_2^1$};
  \node[color=lightgray, state]                                                                      (s3x1)  [below of=C3, label=left:$p_1$]        {$r_3^1$};
  
  \node[state]                                                           (s1x2)  [below of=s1x1, label=left:$p_2$]                                      {$r_1^2$};
  \node[state]                                                           (s2x2)  [below left of=s2x1, label=left:$p_2$]                                         {$r_2^2$};
  \node[color=lightgray, state]                                                              (s3x2)  [below of=s3x1]                                                                                                                {$r_3^2$};
  
  \node[color=lightgray, state]                                                              (s1x3)  [below of=s1x2, label=left:$p_3$]                                      {$r_1^3$};
  \node[state]                                                           (s2x3)  [below of=s2x2]                                                                                                                {$r_2^3$};
  \node[state]                                                                                   (s3x3_) [below left of=s3x2]                                                                                                           {$\overline r_3^3 $};
  \node[state]                                                           (s3x3)  [below right of=s3x2, label=left:$p_3$]                                        {$r_3^3$};  
  \node[state]                                                                   (s1x4)  [below right of=s1x3, label=left:$p_4$]        {$r_1^4$};
  \node[state]                                                                   (s1x4_)  [below left of=s1x3, label=left:$$]           {$\overline r_1^4 $};
  \node[color=lightgray, state]              (s2x4)  [below of=s2x3, label=left:$p_4$]      {$r_2^4$};
  \node[state]           (s3x4)  [below of=s3x3, label=left:$p_4$]      {$r_3^4$}       ;
  \node[state]           (s3x4_)  [below of=s3x3_, label=left:$$]       {$\overline r_3^4 $};

  \path (C1)                    edge            node {}(s1x1)
        (s1x1)                  edge            node {}(s1x2)
        (s1x2)                  edge            node {}(s1x3)
        (s1x3)                  edge            node {}(s1x4_)
                                                edge            node {}(s1x4)
        (C2)                            edge            node {}(s2x1)
                                                edge            node     {}(s2x1_)
        (s2x1)                  edge            node {}(s2x2)
        (s2x1_)                 edge            node {}(s2x2)
        (s2x2)                  edge            node {}(s2x3)
        (s2x3)                  edge            node {}(s2x4)
        (C3)                            edge            node {}(s3x1)
        (s3x1)                  edge            node {}(s3x2)
        (s3x2)                  edge            node    {}(s3x3)
        (s3x2)                  edge            node    {}(s3x3_)
        (s3x3_)                 edge            node {}(s3x4_)
        (s3x3)                  edge            node {}(s3x4);
\end{tikzpicture}
\end{center}
\caption{Kripke structure corresponding to $\varphi\,=\,(\neg x_1 \vee x_2 \vee x_3)\wedge(x_2\vee \neg x_3 \vee x_4)\wedge(x_1 \vee \neg x_2)$}
\label{fig:wedge-vee-np-complete-example}
\end{figure}

Let $T=\{s_1,s_2,s_3\}$ with $W,T \models \psi$ and for all $j \in \{1,\ldots,4\}$ let $T_j \subseteq T$ with $T_j \models \gamma_j$ and $T_1 \cup \ldots \cup T_4 = T$.
By comparing the formulae $\gamma_j$ with the chains in the Kripke structure one can easily verify that $T_1\varsubsetneq \{s_1,s_3\}$ i.e., there can at most be one of $s_1$ and $s_3$ in $T_1$ since $\pi(r_1^1)\cap{p_1} \neq \pi(r_3^1)\cap\{p_1\}$ and $s_2$ cannot be in $T_1$ since its direct successors $\overline r_{2}^1, r_2^1$ do not agree on $p_1$. In this case $T_1 = \{s_1\}$ means that $C_1$ is satisfied by setting $\theta(x_1) = 0$ and the fact that $\{s_2\} \not\models \gamma_1$ corresponds to the fact that there is no way to satisfy $C_2$ via $x_1$, because $x_1$ does not occur in $C_2$.
Analogously, $T_2\subseteq\{s_1,s_2\}$ or $T_2\subseteq\{s_3\}$, and $T_3 \varsubsetneq \{s_1,s_2\}$ and $T_4 \subseteq \{s_2\}$.

Now, e.g., the valuation $\theta$ where $x_1$, $x_3$ and $x_4$ evaluate to true and $x_2$ to false satisfies $\varphi$. From this valuation one can construct sets $T_1, \dots, T_4$ with $T_1\cup\dots\cup T_4=\{s_1,s_2,s_3\}$ such that $T_j \models \gamma_j$ for all $j=1,\dots,4$ by defining $T_j:=\{s_i \mid \text{$x_j$ satisfies clause $C_i$ under }\allowbreak\theta\allowbreak\}$ for all $j$. This leads to $T_1=T_2=\{s_3\}$, $T_3=\{s_1\}$ and $T_4=\{s_2\}$.

The gray colourings indicate which chains (resp.~clauses) are satisfied on which levels (resp.~by which variables).
$\psi$ (resp.~$\varphi$) is satisfied because there is a gray coloured state in each chain.
\end{example}

If we disallow both $\EX$ and $\vee$ the problem becomes tractable since the non-deterministic steps in the model checking algorithm are no longer needed.

\begin{theorem}
\label{box-wedge-in-p}
Let $M \subseteq \{\AX,\wedge,\aneg, \dep{}\}$. Then \MDLMC[M] is in \PTIME.
\end{theorem}
\begin{proof}
Algorithm~\ref{algo:check} is a non-deterministic algorithm that checks the truth of an arbitrary \MDL formula in a given structure in polynomial time. Since $M\subseteq \{\AX,\wedge,\aneg, \dep{}\}$ it holds that $\EX,\,\vee\notin M$. Therefore the non-deterministic steps are never used and the algorithm is in fact deterministic in this case.
\end{proof}

Note that this deterministic polynomial time algorithm is a top-down algorithm and therefore works in a fundamentally different way than the usual deterministic polynomial time bottom-up algorithm for plain modal logic.

Now we have seen that $\MDLMC[M]$ is tractable if $\vee\notin M$ and $\EX \notin M$ since these two operators are the only source of non-determinism.
On the other hand, $\MDLMC[M]$ is \NP-complete if $\dep{}\in M$ and either $\EX\in M$ (Theorem~\ref{diamond-np-complete}) or $\vee,\,\AX\in M$ (Theorem~\ref{box-vee-np-complete}). The remaining question is what happens if only $\vee$ (but not $\AX$) is allowed. Unfortunately this case has to remain open for now.

\section{Bounded arity fragments}\label{sec:bounded-fragments}

We will now show that $\MDLMC[\{\vee, \aneg, \dep{}\}]$ is in \PTIME if we impose the following constraint on the dependence atoms in formulae given as part of problem instances: there is a constant $k\in \N$ such that in any input formula it holds for all dependence atoms of the form $\dep[p_1,\ldots,p_j]{p}$ that $j \leq k$. To prove this statement we will decompose it into two smaller propositions.

First we show that even the whole $\{\vee, \aneg, \dep{}\}$ fragment with unrestricted $\dep{\cdot}$ atoms is in \PTIME as long as it is guaranteed that in every input formula at least a specific number of dependence atoms -- depending on the size of the Kripke structure -- occur.

We will need the following obvious lemma stating that a dependence atom is always satisfied by a team containing at least half of all the worlds.
\begin{lemma}\label{split-in-two}
Let $W=(S,R,\pi)$ be a Kripke structure, $\varphi \dfn\dep[p_1,\ldots,p_n]{q}$ ($n\geq 0$) an atomic formula and $T\subseteq S$ an arbitrary team. Then there is a set $T'\subseteq T$ such that $|T'| \geq \frac{|T|}{2}$ and $T'\models \varphi$.
\end{lemma}
\begin{proof}
Let $T_0:=\{s\in T\mid q\notin\pi(s)\}$ and $T_1:=\{s\in T\mid q\in\pi(s)\}$. Then $T_0\cup T_1=T$ and $T_0\cap T_1=\emptyset$. Therefore there is an $i\in\{1,2\}$ such that $|T_i| \geq \frac{|T|}{2}$. Let $T':=T_i$. Since $q$ is either labeled in every state of $T'$ or in no one, it holds that $W,T'\models \varphi$.
\end{proof}


We will now formalize a notion of \anfz{many dependence atoms in a formula}.

\begin{definition}\label{defMDLMCl}
For $\varphi\in\MDL$ let $\sigma(\varphi)$ be the number of positive dependence atoms in $\varphi$. Let $\ell: \mathbb{N} \to \mathbb{R}$ an arbitrary function and $\star\in\{<,\leq,>,\geq,=\}$.
Then \MDLMCparas[M]{}{\star\,\ell(n)} (resp.~\MDLMCparas[M]{k}{\star \ell(n)}) is the problem \MDLMC[M] (resp.~\MDLMCk[M]) restricted to inputs $\enc{W=(S,R,\pi),T,\varphi}$ that satisfy the condition $\sigma(\varphi)\, \star\, \ell(|S|)$.
\end{definition}

If we only allow $\vee$ and we are guaranteed that there are many dependence atoms in each input formula then model checking becomes trivial -- even for the case of unbounded dependence atoms.

\begin{proposition}\label{many-dep-atoms-trivial}
Let $M\subseteq\{\vee,\aneg,\dep{}\}$. Then $\MDLMCparas[M]{}{> \log_2(n)}$ is trivial, i.e., for all Kripke structures $W=(S,R,\pi)$ and all $\varphi \in \MDL[M]$ such that the number of positive dependence atoms in $\varphi$ is greater than $\log_2(|S|)$ it holds for all $T\subseteq S$ that $W, T \models \varphi$.
\end{proposition}
\begin{proof}
Let $W=(S,R,\pi)$, $\varphi \in \MDL[M]$, $T\subseteq S$ be an arbitrary instance with $\ell > \log_2(|S|)$ dependence atoms in $\varphi$. Then either $\varphi\equiv \true$ or
\[\varphi\ \equiv\ \bigvee_{i=1}^\ell \underbrace{\dep{p_{j_{i,1}},\dots,p_{j_{i,k_i}}}}_{\psi_i}\,\vee\,\bigvee_i l_i,\]
where each $l_i$ is either a (possibly negated) atomic proposition or a negated dependence atom.

{\bfseries Claim.} For all $k\in\{0,\dots, \ell\}$ there is a set $T_k\subseteq T$ such that $W,T_k\models \bigvee_{i=1}^k \psi_i$ and $|T\setminus T_k| < 2^{\ell-k}$.

The main proposition follows immediately from case $k=\ell$ of this claim: From $|T\setminus T_\ell| < 2^{\ell-\ell} = 1$ follows that $T=T_\ell$ and from $W,T_\ell \models \bigvee_{i=1}^\ell \psi_i$ follows that $W,T \models \varphi$.

\emph{Inductive proof of the claim.} For $k=0$ we can choose $T_k := \emptyset$. For the inductive step let the claim be true for all $k'<k$. By Lemma~\ref{split-in-two} there is a set $T'_k\,\subseteq\,T\setminus T_{k-1}$ such that $W,T'_k\models \psi_k$ and $|T'_K| \geq \frac{|T\setminus T_{k-1}|}{2}$. Let $T_k:=T_{k-1}\cup T'_k$. Since $W,T_{k-1}\models \bigvee_{i=1}^{k-1} \psi_i$ it follows by definition of the semantics of $\vee$ that $W,T_k\models \bigvee_{i=1}^{k} \psi_i$. Furthermore,
\[\begin{array}{rcccl}
|T\setminus T_k|&=&|(T\setminus T_{k-1}) \setminus T'_k| & = & |T\setminus T_{k-1}| - |T'_k|\\
&\leq& |T\setminus T_{k-1}| - \frac{|T\setminus T_{k-1}|}{2} &=& \frac{|T\setminus T_{k-1}|}{2}\\
&<& \frac{2^{\ell-(k-1)}}{2} &=& 2^{\ell-k}.
\end{array}\]
\end{proof}
Note that $\MDLMCparas[M]{}{> \log_2(n)}$ is only trivial, i.e., all instance structures satisfy all instance formulae, if we assume that only valid instances, i.e., where the number of dependence atoms is guaranteed to be large enough, are given as input. However, if we have to verify this number the problem clearly remains in \PTIME.

Now we consider the case in which we have very few dependence atoms (which have bounded arity) in each formula. We use the fact that there are only a few dependence atoms by searching through all possible determinating functions for the dependence atoms. Note that in this case we do not need to restrict the set of allowed \MDL operators as we have done above.

\begin{proposition}
\label{few-dep-atoms-in-p}
Let $k\geq 0$. Then $\MDLMCparas{k}{\leq \log_2(n)}$ is in \PTIME.
\end{proposition}
\begin{proof}
From the semantics of $\dep{}$ it follows that $\dep[p_1,\dots,p_k]{p}$ is equivalent to
\begin{equation}\label{eq:dep-as-function}
\exists f\,f(p_1,\dots,p_{k})\leftrightarrow p\quad :=\quad \exists f\,\big( (\neg f(p_1,\dots,p_k) \vee  p) \wedge (f(p_1,\dots,p_k) \vee \neg p) \big)
\end{equation}
where $f(p_1,\dots,p_k)$ and $\exists f \varphi$ -- both introduced by Sevenster \cite[Section~4.2]{se09} -- have the following semantics:
\[\begin{array}{l@{\quad\text{iff}\quad}p{7cm}}
W,T\models \exists f \varphi            & there is a Boolean function $f^W$ such that $(W,f^W),\allowbreak T\models \varphi$\\
(W,f^W),T\models f(p_1,\dots,p_k)       & for all $s\in T$ and for all $x_1,\dots,x_k\in\{0,1\}$ with $x_i=1$ iff $p_i\in\pi(s)$ $(i=1,\dots,k)$: $f^W(x_1,\dots,x_k)=1$\\
(W,f^W),T\models \neg f(p_1,\dots,p_k)  & for all $s\in T$ and for all $x_1,\dots,x_k\in\{0,1\}$ with $x_i=1$ iff $p_i\in\pi(s)$ $(i=1,\dots,k)$: $f^W(x_1,\dots,x_k)=0$\\
\end{array}\]

Now let $W=(S,R,\pi)$, $T\subseteq S$ and $\varphi\in \MDL_k$ be a problem instance. First, we count the number $\ell$ of dependence atoms in $\varphi$. If $\ell > \log_2(|S|)$ we reject the input instance.
Otherwise we replace every dependence atom by its translation according to \ref{eq:dep-as-function} (each time using a new function symbol).
Since the dependence atoms in $\varphi$ are at most $k$-ary we have from the transformation \eqref{eq:dep-as-function} that the introduced function variables $f_1,\dots,f_\ell$ are also at most $k$-ary. From this it follows that the upper bound for the number of interpretations of each of them is $2^{2^k}$.
For each possible tuple of interpretations $f_1^W,\dots,f_\ell^W$ for the function variables we obtain an \ML formula $\varphi^*$ by replacing each existential quantifier $\exists f_i$ by a Boolean formula encoding of the interpretation $f_i^W$ (for example by encoding the truth table of $f_i$ with a formula in disjunctive normal form).
For each such tuple we model check $\varphi^*$. That is possible in polynomial time in $|S| + |\varphi^*|$ as shown by Clarke et al.~\cite{clemsi86}. Since the encoding of an arbitrary $k$-ary Boolean function has length at most $2^{k}$ and $k$ is constant this is a polynomial in $|S| + |\varphi|$.

Furthermore, the number of tuples over which we have to iterate is bounded by
\[\begin{array}{rclcl}
\left(2^{2^k}\right)^{\log_2(|S|) } & = & 2^{2^k\cdot \log_2(|S|)}\\
 & = & \left(2^{\log_2(|S|)}\right)^{2^k}\\
 & = & |S|^{2^k} &\in & |S|^{\mathCommandFont{O}(1)}.
\end{array}\]
\end{proof}

With Proposition~\ref{many-dep-atoms-trivial} and Proposition~\ref{few-dep-atoms-in-p} we have shown the following theorem.
\begin{theorem}
\label{vee-bounded-in-p}
Let $M\subseteq \{\vee,\aneg,\dep{}\}$, $k\geq 0$. Then \MDLMCk[M] is in \PTIME.
\end{theorem}
\begin{proof}
Given a Kripke structure $W=(S,R,\pi)$ and a $\MDL_k(\vee, \aneg,\dep{})$ formula $\varphi$ the algorithm counts the number $m$ of dependence atoms in $\varphi$. If $m> \log_2(|S|)$ the input is accepted (because by Proposition~\ref{many-dep-atoms-trivial} the formula is always fulfilled in this case). Otherwise the algorithm from the proof of Proposition~\ref{few-dep-atoms-in-p} is used.
\end{proof}

And there is another case where we can use the exhaustive determinating function search.
\begin{theorem}
\label{no-wedge-no-vee}
Let $M\subseteq\{\AX, \EX, \aneg, \dep{}\}$. Then $\MDLMCk[M]$ is in $\PTIME$ for every $k\geq 0$.
\end{theorem}
\begin{proof}
Let $\varphi\in \MDL_k(M)$. Then there can be at most one dependence atom in $\varphi$ because $M$ only contains unary operators. Therefore we can once again use the algorithm from the proof of Proposition~\ref{few-dep-atoms-in-p}.
\end{proof}

In Theorem~\ref{diamond-np-complete} we saw that $\MDLMC[\EX, \dep{}]$ is \NP-complete.
The previous theorem includes $\MDLMCk[\EX,\dep{}] \in \PTIME$ as a special case.
Hence, the question remains which are the minimal supersets $M$ of $\{\EX,\dep{}\}$ such that $\MDLMCk[M]$ is \NP-complete.

We will now see that adding either $\wedge$ (\Cref{diamond-wedge-bounded}) or $\vee$ (\Cref{diamond-vee-bounded}) is already enough to get \NP-completeness again.
But note that in the case of $\vee$ we need $k\geq 1$ while for $k=0$ the question remains open.
\begin{theorem}\label{diamond-wedge-bounded}
Let $M\supseteq\{\EX, \wedge, \dep{}\}$. Then $\MDLMCk[M]$ is \NP-complete for every $k\geq 0$.
\end{theorem}
\begin{proof}
Membership in \NP follows from \cref{mdlmc-in-np}. For hardness we once again reduce \ThreeSAT to our problem.

For this purpose let $\varphi \dfn \bigwedge_{i=1}^m C_i$ be an arbitrary \cnf formula built from the variables $x_1, \dots, x_n$. Let $W$ be the Kripke structure $(S, R, \pi)$ shown in \cref{figure:diamond-wedge-bounded} and formally defined by
\[\begin{array}{lcl}
S & \dfn & \{c_i \mid i\in\{1,\dots, m\}\} \cup \{s_{j,j'}, \overline{s}_{j,j'} \mid j,j'\in\{1,\dots,n \}\}\\
  &    & \cup\ \{t_j,\overline{t}_j\mid j\in\{1,\dots,n\}\}\\
R & \dfn & \{(c_i, s_{1,j}) \mid x_j \in C_i\} \cup \{(c_i, \overline{s}_{1,j}) \mid \overline x_j  \in C_i\}\\
  &    & \cup \ \{(s_{k,j}, s_{k+1,j}) \mid j\in \{1,\dots,n\}, k \in \{1,\dots, n-1\}\}\\
  &    & \cup \ \{(\overline{s}_{k,j}, \overline{s}_{k+1,j}) \mid j\in \{1,\dots,n\}, k \in \{1,\dots, n-1\}\}\\
  &    & \cup \ \{(s_{k,j}, t_j), (\overline{s}_{k,j},\overline{t}_j) \mid j\in\{1,\dots,n\}, k\in \{1,\dots,n\}\}\\
  &    & \cup \ \{(s_{k,j}, \overline{t}_j), (\overline{s}_{k,j}, t_j) \mid j\in\{1,\dots,n\}, k\in \{1,\dots,n\}, j\neq k\}\\
\pi(c_i) & \dfn & \emptyset\\
\pi(s_{j,j'}) & \dfn & \emptyset
\\
\pi(\overline{s}_{j,j'}) & \dfn & \emptyset
\\
\pi(t_j) & \dfn & \{r_j, p_j\}\\
\pi(\overline{t}_j) & \dfn & \{r_j\}.\\
\end{array}\]

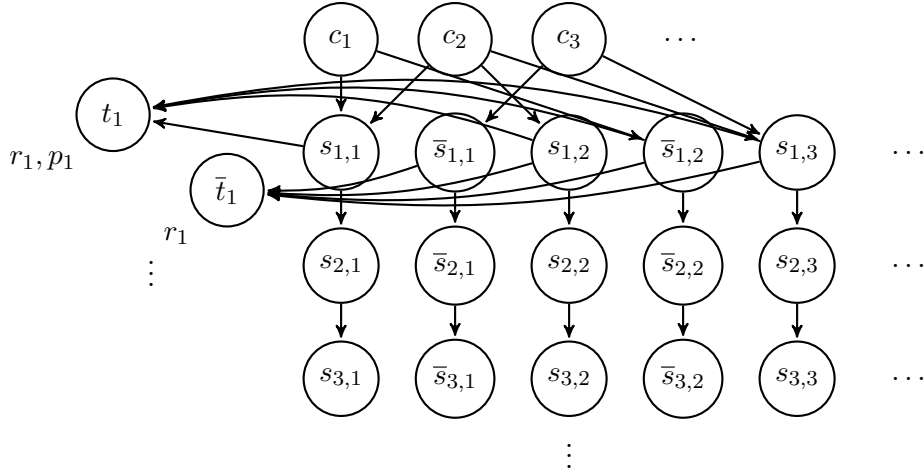
\begin{figure}[ht]
\begin{center}
\begin{tikzpicture}[->,>=stealth',shorten >=1pt,auto,node distance=1.5cm,thick]
\node[state]                             (C1)                      {$c_1$};
\node[state]                             (C2)       [right of=C1]               {$c_2$};
\node[state]                             (C3)       [right of=C2]               {$c_3$};
\node                            (Cx)       [right of=C3]               {$\dots$};

\node[state]                             (x1)       [below of=C1]               {$s_{1,1}$};
\node[state]                             (tx1)       [right of=x1]               {$\overline{s}_{1,1}$};
\node[state]                             (x2)       [right of=tx1]               {$s_{1,2}$};
\node[state]                             (tx2)       [right of=x2]               {$\overline{s}_{1,2}$};
\node[state]                             (x3)       [right of=tx2]               {$s_{1,3}$};
\node                            (xx)       [right of=x3]               {$\dots$};
\node[state]                             (tt1)       at  (-1.5,-2)  [label=225:{$r_1$}]               {$\overline{t}_{1}$};
\node[state]                             (t1)        at  (-3,-1)    [label=225:{$r_1,p_1$}]           {$t_{1}$};

\node[state]                             (xx1)       [below of=x1]               {$s_{2,1}$};
\node                            (xx)       at (-2.5,-3)               {$\vdots$};
\node[state]                             (txx1)       [below of=tx1]               {$\overline{s}_{2,1}$};
\node[state]                             (xx2)       [below of=x2]               {$s_{2,2}$};
\node[state]                             (txx2)       [below of=tx2]              {$\overline{s}_{2,2}$};
\node[state]                             (xx3)       [below of=x3]               {$s_{2,3}$};
\node                            (xx)       [right of=xx3]               {$\dots$};

\node[state]                             (xxx1)       [below of=xx1]               {$s_{3,1}$};
\node[state]                             (txxx1)       [below of=txx1]               {$\overline{s}_{3,1}$};
\node[state]                             (xxx2)       [below of=xx2,label=below:{$\vdots$}]               {$s_{3,2}$};
\node[state]                             (txxx2)       [below of=txx2]              {$\overline{s}_{3,2}$};
\node[state]                             (xxx3)       [below of=xx3]               {$s_{3,3}$};
\node                            (xxx)       [right of=xxx3]               {$\dots$};


\path (C1) edge  node {}(x1);
\path (C1) edge  node {}(tx2);
\path (C2) edge  node {}(x1);
\path (C2) edge  node {}(x2);
\path (C2) edge  node {}(x3);
\path (C3) edge  node {}(tx1);
\path (C3) edge  node {}(x3);

\path (x1) edge  node {}(xx1);
\path (tx1) edge  node {}(txx1);
\path (x2) edge  node {}(xx2);
\path (tx2) edge  node {}(txx2);
\path (x3) edge  node {}(xx3);
\path (x1) edge              node {}(t1);
\path (x2) edge [bend right=14]  node {}(t1);
\path (tx2) edge [bend right=14]  node {}(t1);
\path (x3) edge [bend right=14]  node {}(t1);
\path (tx1) edge [bend left=10]  node {}(tt1);
\path (x2) edge [bend left=10]  node {}(tt1);
\path (tx2) edge [bend left=10]  node {}(tt1);
\path (x3) edge [bend left=10]  node {}(tt1);

\path (xx1) edge  node {}(xxx1);
\path (txx1) edge  node {}(txxx1);
\path (xx2) edge  node {}(xxx2);
\path (txx2) edge  node {}(txxx2);
\path (xx3) edge  node {}(xxx3);
\end{tikzpicture}
\caption{Kripke structure construction in the proof of \cref{diamond-wedge-bounded}}
{\small The underlying \cnf formula contains the clauses $C_1=x_1\vee\neg x_2$, $C_2=x_1 \vee x_2\vee x_3$ and $C_3=\neg x_1\vee x_3$}
\label{figure:diamond-wedge-bounded}
\end{center}
\end{figure}

And let $\psi$ be the \MDL[\EX,\wedge,\dep{}] formula
\[\begin{array}{cl}
& \EX\left(\bigwedge\limits_{j=1}^n\, \EX^{j} (r_j \wedge \dep{p_j}) \right)\\
= & \EX\big(\EX(r_1\wedge \dep{p_1})\,\wedge\, \EX\EX (r_2\wedge\dep{p_2}) \,\wedge\, \dots\, \wedge\, \EX^{n}(r_n\wedge \dep{p_n})\big).
\end{array}\]

We again show that $\varphi \in \ThreeSAT$ iff $W,\{c_1,\dots, c_m\} \models \psi$.
First assume that $\varphi \in \ThreeSAT$ and that $\theta$ is a satisfying valuation for the variables in $\varphi$.
Now let
\[s_j:=\begin{cases}
               s_{1,j} & \text{if $x_j$ evaluates to true under $\theta$}\\
               \overline s_{1,j} & \text{if $x_j$ evaluates to false under $\theta$}
              \end{cases}\]
for all $j=1,\dots,n$.
Then it holds that $W, \{s_1,\dots,s_n\} \models \bigwedge\limits_{j=1}^n\, \EX^{j} (r_j \wedge \dep{p_j})$.

Furthermore, since $\theta$ satisfies $\varphi$ it holds for all $i=1,\dots,m$ that there is a $j_i\in\{1,\dots,n\}$ such that $(c_i,s_{j_i})\in R$.
Hence, $W, \{c_1,\dots,c_m\} \models \EX\left(\bigwedge\limits_{j=1}^n\, \EX^{j} (r_j \wedge \dep{p_j})\right)$.

%

For the reverse direction assume that $W, \{c_1,\dots,c_m\}  \models \psi$.
Now let $T\subseteq \{s_{1,1},\allowbreak\overline s_{1,1},\allowbreak s_{1,2},\allowbreak\dots,\overline s_{1,n}\}$ such that $T\models \bigwedge\limits_{j=1}^n\, \EX^{j} (r_j \wedge \dep{p_j})$ and for all $i=1,\dots,m$ there is a $s\in T$ with $(c_i,s)\in R$.

Since $T\models \EX^{j} (r_j \wedge \dep{p_j})$ there is no $j\in\{1,\dots,n\}$ with $s_{1,j}\in T$ and also $\overline s_{1,j}\in T$.
Now let $\theta$ be the valuation of $x_1,\dots,x_n$ defined by
\[\theta(x_j):=\begin{cases}
  1 & \text{if $s_{1,j}\in T$}\\
  0 & \text{else}.
\end{cases}\]

Since for each $i=1,\dots,m$ there is a $j\in\{1,\dots,n\}$ such that either $(c_{i},s_{1,j})\in R$ and $s_{1,j}\in T$ or $(c_{i},\overline s_{1,j})\in R$ and $\overline s_{1,j}\in T$ it follows that for each clause $C_i$ of $\varphi$ there is a $j\in\{1,\dots,n\}$ such that $x_j$ satisfies $C_i$ under $\theta$.
\end{proof}

\begin{theorem}
\label{diamond-vee-bounded}
Let $M\supseteq\{\EX, \vee, \dep{}\}$. Then $\MDLMCk[M]$ is \NP-complete for every $k\geq 1$.
\end{theorem}
\begin{proof}
As above membership in \NP follows from Proposition \ref{mdlmc-in-np} and for hardness we reduce \ThreeSAT to our problem.

For this purpose let $\varphi \dfn \bigwedge_{i=1}^m C_i$ be an arbitrary \CNF formula built from the variables $p_1, \dots, p_n$. Let $W$ be the Kripke structure $(S, R, \pi)$ shown in Figure \ref{figure:diamond-vee-bounded} and formally defined by
\[\begin{array}{lcl}
S & \dfn & \{c_{i,j} \mid i\in\{1,\dots, m\}, j\in\{1,\dots,n\}\} \cup \{x_{j,j'} \mid j,j'\in\{1,\dots,n \}, j'\leq j\}\\
R & \dfn & \{(c_{i,j}, c_{i,j+1})\mid i\in\{1,\dots, m\}, j\in\{1,\dots, n-1\}\}\\
  &    & \cup\  \{(x_{j,j'},x_{j,j'+1}) \mid j\in\{1,\dots,n\}, j'\in\{1,\dots, j-1\}\}\\
\pi(x_{j,j'}) & \dfn & \left \{ \begin{array}{ll}
  \{q, p_j\} & \text{iff } j'=j\\
  \{q\} & \text{iff } j'<j
\end{array}\right.\\
\pi(c_{i,j}) & \dfn & \left \{ \begin{array}{lll}
  \{q\} & \text{iff } p_j, \neg p_j \notin C_i\\[0.5ex]
  \{p_j\} & \text{iff } p_j\in C_i\\
  \emptyset & \text{iff } \neg p_j \in C_i
\end{array}\right.
\end{array}\]

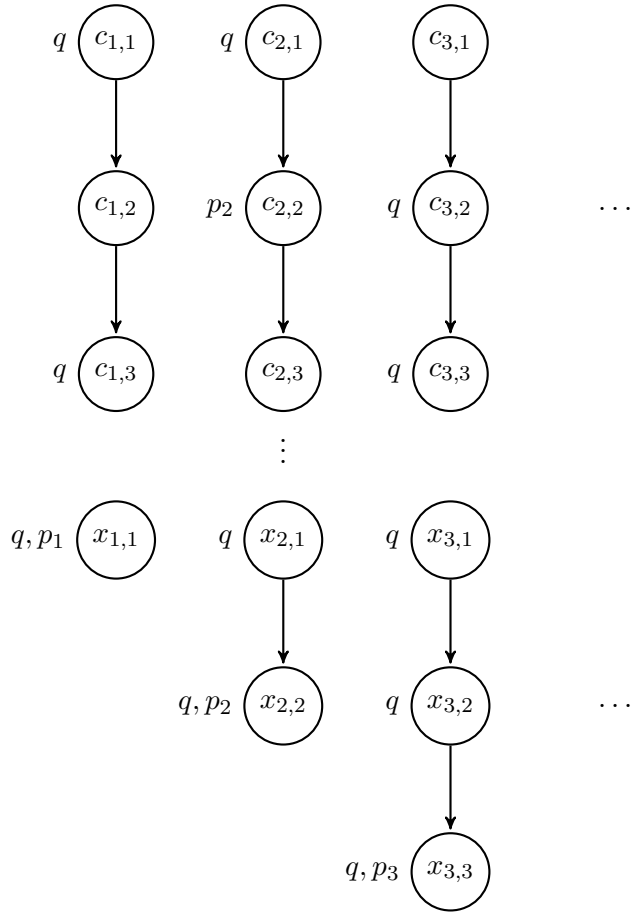
\begin{figure}[ht]
\begin{center}
\begin{tikzpicture}[->,>=stealth',shorten >=1pt,auto,node distance=2.2cm,thick]
\node[state]                             (c1)       [label=left:{$q$}]               {$c_{1,1}$};
\node[state]                             (c2)       [right of=c1,label=left:{$q$}]               {$c_{2,1}$};
\node[state]                             (c3)       [right of=c2]               {$c_{3,1}$};
\node[state]                             (c12)      [below of=c1]                {$c_{1,2}$};
\node[state]                             (c22)      [below of=c2,label=left:{$p_2$}]                {$c_{2,2}$};
\node[state]                             (c32)      [below of=c3,label=left:{$q$}]                {$c_{3,2}$};
\node[state]                             (c13)      [below of=c12,label=left:{$q$}]                {$c_{1,3}$};
\node[state]                             (c23)      [below of=c22,label=below:{$\vdots$}]                {$c_{2,3}$};
\node[state]                             (c33)      [below of=c32,label=left:{$q$}]                {$c_{3,3}$};
\node                            (cx)       [right of=c32]               {$\dots$};

\node[state]                             (x2)       [below of=c23,label=left:{$q$}]               {$x_{2,1}$};
\node[state]                             (x1)       [left of=x2,label=left:{$q,p_1$}]               {$x_{1,1}$};
\node[state]                             (x3)       [right of=x2,label=left:{$q$}]               {$x_{3,1}$};

\node[state]                             (xx2)       [below of=x2,label=left:{$q,p_2$}]               {$x_{2,2}$};
\node[state]                             (xx3)       [below of=x3,label=left:{$q$}]               {$x_{3,2}$};
\node                            (xx)       [right of=xx3]               {$\dots$};

\node[state]                             (xxx3)       [below of=xx3,label=left:{$q,p_3$}]               {$x_{3,3}$};

\path (c1) edge  node {}(c12);
\path (c2) edge  node {}(c22);
\path (c3) edge  node {}(c32);
\path (c12) edge  node {}(c13);
\path (c22) edge  node {}(c23);
\path (c32) edge  node {}(c33);

\path (x2) edge  node {}(xx2);
\path (x3) edge  node {}(xx3);

\path (xx3) edge  node {}(xxx3);
\end{tikzpicture}
\caption{Kripke structure construction in the proof of Theorem \ref{diamond-vee-bounded}\newline
\small{The underlying \cnf formula contains the clauses $C_1=\neg p_2$, $C_2=p_2\vee \neg p_3$ and $C_3=\neg p_1$}}
\label{figure:diamond-vee-bounded}
\end{center}
\end{figure}

Let $\psi$ be the \MDL formula
\[\begin{array}{cl}
  & \bigvee\limits_{j=1}^n \EX^{j-1} \dep[q]{p_j}\\
\equiv & \dep[q]{p_1} \vee \EX \dep[q]{p_2} \vee \EX \EX \dep[q]{p_3} \vee \dots \vee \EX^{n-1}\dep[q]{p_n}.
\end{array}\]

Once again we show that $\varphi \in \ThreeSAT$ iff $W, \{c_{1,1},\dots, c_{m,1},\, x_{1,1}, x_{2,1},\dots, x_{n,1}\}\models \psi$.
First assume that $\varphi \in \ThreeSAT$ and that $\theta$ is a satisfying valuation for the variables in $\varphi$.
Now let $P_j:=\{c_{i,1}\mid \text{$C_i$ is satisfied by $p_j$ under $\theta$}\}$ for all $j=1,\dots,n$. Then it follows that $\bigcup\limits_{j=1}^n P_j = \{c_{1,1},\dots,c_{m,1}\}$ and that
\[W,P_j \models \EX^{j-1} (\neg q \wedge \dep{p_j})\]
for all $j=1,\dots,n$.
Additionally, it holds that $W,\{x_{j,1}\}\models \EX^{j-1} (q \wedge \dep{p_j})$ ($j=1,\dots,n$).

Together it follows that $W,P_j\cup\{x_{j,1}\}\models \EX^{j-1} \dep[q]{p_j}$ for all $j=1,\dots,n$. This implies
\[W,\bigcup_{j=1}^n (P_j\cup \{x_{j,1}\}) \models \bigvee_{j=1}^n \EX^{j-1} \dep[q]{p_j}\]
which is equivalent to
\[W,\{c_{1,1},\dots, c_{m,1},\, x_{1,1}, x_{2,1},\dots, x_{n,1}\} \models \psi.\]

For the reverse direction assume that $W,T \models \psi$ with $T:=\{c_{1,1},\dots, c_{m,1},\allowbreak\, x_{1,1},\allowbreak x_{2,1},\allowbreak\dots, x_{n,1}\}$.
Let $T_1,\ldots,T_n$ be subsets of $T$ with $T_1 \cup \dots \cup T_n = T$ such that for all $j\in\{1,\dots,n\}$ it holds that $T_j\models \EX^{j-1} \dep[q]{p_j}$.
Then it follows that $x_{1,1}\in T_1$ since the chain starting in $x_{1,1}$ consists of only one state. From $\pi(x_{1,1})=\{q,p_1\}$ and $\pi(x_{2,1})=\{q\}$ it follows that $x_{2,1}\notin T_1$ and hence (again because of the length of the chain) $x_{2,1}\in T_2$. Inductively, it follows that $x_{j,1}\in T_j$ for all $j=1,\dots,n$.

Now, it follows from $x_{j,1}\in T_j$ that for all $i\in\{1,\dots,m\}$ with $c_{i,1}\in T_j$: $q\notin \pi(c_{i,j})$ (because $q,p_j\in\pi(x_{j,j}$, $p_j\notin \pi(x_{i,j})$). Since $T_j\models \EX^{j-1} \dep[q]{p_j}$, it then holds that $T_j\setminus\{x_{j,1}\}\models \EX^{j-1} (\neg q \wedge \dep{p_j})$.

Now let $\theta$ be the valuation of $p_1,\dots,p_n$ defined by
\[\theta(p_j):=\begin{cases}
  1 & \text{if $T_j\setminus\{x_{j,1}\}\models \EX^{j-1} (\neg q \wedge p_j)$}\\
  0 & \text{if $T_j\setminus\{x_{j,1}\}\models \EX^{j-1} (\neg q \wedge \neg p_j)$}
\end{cases}.\]

Since for each $i=1,\dots,m$ there is a $j\in\{1,\dots,n\}$ such that $c_{i,1}\in T_j$ it follows that for each clause $C_i$ of $\varphi$ there is a $j\in\{1,\dots,n\}$ such that $p_j$ satisfies $C_i$ under $\theta$.
\end{proof}

\section{Classical disjunction}\label{sec:mdl-mc-nor}

First we show that classical disjunction can substitute zero-ary dependence atoms.

\begin{lemma}\label{nor replaces dep}
Let $\dep{},\nor \notin M$. Then
\[\MDLMCpara[{M\cup\{\dep{}\}}]{0} \leqpm \MDLMC[{M \cup \{\nor\}}].\]
\end{lemma}
\begin{proof}
Follows immediately from the equivalence of $\dep{p}$ and $p\nor \neg p$ together with \cref{neg-dont-matter}.
\end{proof}

The following surprising result shows that both kinds of disjunctions together are already enough to get \NP-completeness.

\begin{theorem}\label{nor vee}
Let $\{\sor, \nor\}\subseteq M$. Then $\MDLMCk[M]$ is \NP-complete for every $k\geq 0$.
\end{theorem}
\begin{proof}
As above membership in \NP follows from \cref{mdlmc-in-np} and for hardness we reduce \ThreeSAT to our problem -- using a construction that bears some similarities with the one used in the proof of \cref{diamond-vee-bounded}.

For this purpose let $\varphi \dfn \bigwedge_{i=1}^m C_i$ be an arbitrary \cnf formula built from the variables $p_1, \dots, p_n$. Let $W$ be the Kripke structure $(S, R, \pi)$ shown in \cref{figure:nor vee} and formally defined by
\[\begin{array}{lcl}
S & \dfn & \{c_{i} \mid i\in\{1,\dots, m\}\}\\
R & \dfn & \emptyset\\
\pi(c_{i}) & \dfn & \{p_j \mid p_j \in C_i\} \cup \{q_j \mid \neg p_j \in C_i\}.
\end{array}\]

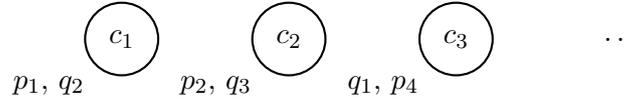
\begin{figure}[ht]
\begin{center}
\begin{tikzpicture}[->,>=stealth',shorten >=1pt,auto,node distance=2.2cm,thick]
\node[state]                             (c1)       [label=225:{$p_1$, $q_2$}]               {$c_{1}$};
\node[state]                             (c2)       [right of=c1,label=225:{$p_2$, $q_3$}]               {$c_{2}$};
\node[state]                             (c3)       [right of=c2,label=225:{$q_1$, $p_4$}]               {$c_{3}$};
\node                            (cx)       [right of=c3]               {$\dots$};
\end{tikzpicture}
\caption{Kripke structure construction in the proof of \cref{nor vee}}
{\small The underlying \cnf formula contains the clauses $C_1=p_1\vee \neg p_2$, $C_2=p_2\vee \neg p_3$ and $C_3=\neg p_1\vee p_4$}
\label{figure:nor vee}
\end{center}
\end{figure}

Let $\psi$ be the \MDL formula
\[\bigvee\limits_{j=1}^n (p_j \nor q_j).\]

Once again we show that $\varphi \in \ThreeSAT$ iff $W, \{c_{1},\dots, c_{m}\}\models \psi$.
First assume that $\varphi \in \ThreeSAT$ and that $\theta$ is a satisfying valuation for $\varphi$.
Now let
\[P_j:=\{c_{i}\mid \text{$C_i$ is satisfied by $p_j$ under $\theta$}\}\]
for all $j=1,\dots,n$. Then it follows that $\bigcup_{j=1}^n P_j = \{c_{1},\dots,c_{m}\}$ and that
\[W,P_j \models p_j \nor q_j\]
for all $j=1,\dots,n$.
Together it follows that
\[W,\{c_{1},\dots, c_{m}\} \models \bigvee\limits_{j=1}^n (p_j \nor q_j).\]

For the reverse direction assume that $W,T \models \psi$ with $T:=\{c_{1},\dots, c_{m}\}$.
Let $T_1,\dots,T_n$ be subsets of $T$ with $T_1 \cup \dots \cup T_n = T$ such that for all $j\in\{1,\dots,n\}$ it holds that $T_j\models p_j\nor q_j$.
Now let $\theta$ be the valuation of $p_1,\dots,p_n$ defined by
\[\theta(p_j):=\begin{cases}
  1 & \text{if }T_j\models p_j\\
  0 & \text{if }T_j\models q_j\\
\end{cases}.\]

Since for each $i=1,\dots,m$ there is a $j\in\{1,\dots,n\}$ such that $c_{i}\in T_j$ it follows that for each clause $C_i$ of $\varphi$ there is a $j\in\{1,\dots,n\}$ such that $p_j$ satisfies $C_i$ under $\theta$.
\end{proof}

Now we show that \cref{no-wedge-no-vee} still holds if we additionally allow classical disjunction.
\begin{theorem}\label{nor and unary}
Let $M\subseteq \{\AX,\EX, \nor, \neg, \dep{}\}$. Then $\MDLMCk[M]$ is in \PTIME for every $k\geq 0$.
\end{theorem}
\begin{proof}
Let $\varphi \in \MDL[M]$.
Because of the distributivity of $\nor$ with all other \MDL operators there is a formula $\psi$ equivalent to $\varphi$ which is of the form
\[\bignor_{i=1}^{|\varphi|} \psi_i\]
with $\psi_i \in \MDL[{M\setminus \{\nor\}}]$ for all $i\in\{1,\dots,|\varphi|\}$.
Note that there are only linearly many formulas $\psi_i$ because $\varphi$ does not contain any binary operators aside from $\nor$.
Further note that $\psi$ can be easily computed from $\varphi$ in polynomial time.

Now it is easy to check for a given structure $W$ and team $T$ whether $W,T\models \psi$ by simply checking whether $W,T\models \psi_i$ (which can be done in polynomial time by \cref{no-wedge-no-vee}) consecutively for all $i\in\{1,\dots,|\varphi|\}$.
\end{proof}

\section{Conclusion}
In this paper we showed that \MDLMC is \NP-complete (Theorem~\ref{wedge-vee-np-complete}). Furthermore we have systematically analyzed the complexity of model checking for fragments of \MDL defined by restricting the set of modal and propositional operators. It turned out that there are several fragments which stay \NP-complete, e.g., the fragment obtained by restricting the set of operators to only $\AX, \vee$ and $\dep{}$ (Theorem~\ref{box-vee-np-complete}) or only $\EX$ and $\dep{}$ (Theorem~\ref{diamond-np-complete}). Intuitively, in the former case the \NP-hardness arises from existentially guessing partitions of teams while evaluating disjunctions and in the latter from existentially guessing successor teams while evaluating $\EX$ operators.
Consequently, if we allow all operators except $\EX$ and $\vee$ the complexity drops to \PTIME (Theorem~\ref{box-wedge-in-p}).


For the fragment only containing $\vee$ and $\dep{}$ on the other hand we were not able to determine whether its model checking problem is tractable.
Our inability to prove either \NP-hardness or containment in \PTIME led us to restrict the arity of the dependence atoms. For the aforementioned fragment the complexity drops to $\PTIME$ in the case of bounded arity (Theorem~\ref{diamond-vee-bounded}). Furthermore, some of the cases which are known to be \NP-complete for the unbounded case drop to \PTIME in the bounded arity case as well (Theorem~\ref{no-wedge-no-vee}) while others remain \NP-complete but require a new proof technique (Theorems~\ref{diamond-wedge-bounded} and~\ref{diamond-vee-bounded}).
Most noteworthy in this context are probably the results concerning the $\EX$ operator. With unbounded dependence atoms this operator alone suffices to get \NP-completeness whereas with bounded dependence atoms it needs the additional expressiveness of either $\wedge$ or $\vee$ to get \NP-hardness.

Considering the classical disjunction operator $\nor$, we showed that the complexity of $\MDLMCpara[{M\cup \{\dep{}\}}]{k}$ is never higher than the complexity of $\MDLMCk[{M\cup \{\nor\}}]$, i.e., $\nor$ is at least as bad as $\dep{\cdot}$ with respect to the complexity of model-checking (in contrast to the complexity of satisfiability; cf.~\cite{lovo10}). And in the case where only $\vee$ is allowed we even have a higher complexity with $\nor$ (\Cref{nor vee}) than with $\dep{}$ (\Cref{vee-bounded-in-p}). The case of $\MDLMC[\vee,\nor]$ is also our probably most surprising result since the non-determinism of the $\vee$ operator turned out to be powerful enough to lead to \NP-completeness although neither conjunction nor dependence atoms (which also, in a sense, contain some special kind of conjunction) are allowed.

Interestingly, in none of our reductions to show \NP-hardness the \MDL formula depends on anything else but the number of propositional variables of the input \cnf formula. The structure of the input formula is always encoded by the Kripke structure alone.
So it seems that even for a fixed formula the model checking problem could still be hard.
This, however, cannot be the case since, by Theorem~\ref{few-dep-atoms-in-p}, model checking for a fixed formula is always in \PTIME. 

Another open question, apart from the unclassified unbounded arity case, is related to a case with bounded arity dependence atoms. In \cref{diamond-vee-bounded} it was only possible to prove \NP-hardness for arity at least one and it is not known what happens in the case where the arity is zero.
Additionally, it might be interesting to determine the exact complexity for the cases which are in \PTIME since we have not shown any lower bounds in these cases so far.



\begin{thebibliography}{BMM{\etalchar{+}}11}

\bibitem[AV09]{abva08}
Samson Abramsky and Jouko V{\"a}{\"a}n{\"a}nen.
\newblock From {IF} to {BI}.
\newblock {\em Synthese}, 167(2):207--230, 2009.

\bibitem[BMM{\etalchar{+}}11]{bememuscthvo11}
Olaf Beyersdorff, Arne Meier, Martin Mundhenk, Thomas Schneider, Michael
  Thomas, and Heribert Vollmer.
\newblock Model checking {CTL} is almost always inherently sequential.
\newblock {\em Logical Methods in Computer Science}, 2011.

\bibitem[CES86]{clemsi86}
E.~M. Clarke, E.~A. Emerson, and A.~P. Sistla.
\newblock Automatic verification of finite-state concurrent systems using
  temporal logic specifications.
\newblock {\em ACM Trans. Program. Lang. Syst.}, 8(2):244--263, 1986.

\bibitem[EL12]{eblo12}
Johannes Ebbing and Peter Lohmann.
\newblock Complexity of model checking for modal dependence logic.
\newblock In {\em Proceedings SOFSEM 2012: Theory and Practice of Computer
  Science}, volume 7147 of {\em Lecture Notes in Computer Science}, pages
  226--237. Springer Berlin / Heidelberg, 2012.

\bibitem[Hem05]{he05}
Edith Hemaspaandra.
\newblock The complexity of poor man's logic.
\newblock {\em CoRR}, cs.LO/9911014v2, 2005.

\bibitem[HSS10]{hescsc10}
Edith Hemaspaandra, Henning Schnoor, and Ilka Schnoor.
\newblock Generalized modal satisfiability.
\newblock {\em J. Comput. Syst. Sci.}, 76(7):561--578, 2010.

\bibitem[Lew79]{le79}
Harry Lewis.
\newblock Satisfiability problems for propositional calculi.
\newblock {\em Mathematical Systems Theory}, 13:45--53, 1979.

\bibitem[LV10]{lovo10}
Peter Lohmann and Heribert Vollmer.
\newblock Complexity results for modal dependence logic.
\newblock In {\em Proceedings 19th Conference on Computer Science Logic},
  volume 6247 of {\em Lecture Notes in Computer Science}, pages 411--425.
  Springer Berlin / Heidelberg, 2010.

\bibitem[Sev09]{se09}
Merlijn Sevenster.
\newblock Model-theoretic and computational properties of modal dependence
  logic.
\newblock {\em Journal of Logic and Computation}, 19(6):1157--1173, 2009.

\bibitem[V{\"a}{\"a}07]{va07}
Jouko V{\"a}{\"a}n{\"a}nen.
\newblock {\em Dependence logic: A new approach to independence friendly
  logic}.
\newblock Number~70 in London Mathematical Society student texts. Cambridge
  University Press, 2007.

\bibitem[V{\"a}{\"a}08]{va08}
Jouko V{\"a}{\"a}n{\"a}nen.
\newblock Modal dependence logic.
\newblock In Krzysztof~R. Apt and Robert van Rooij, editors, {\em New
  Perspectives on Games and Interaction}, volume~4 of {\em Texts in Logic and
  Games}, pages 237--254. Amsterdam University Press, 2008.

\end{thebibliography}
\newcommand{\etalchar}[1]{$^{#1}$}

\addcontentsline{toc}{section}{References}

\end{document}